\documentclass[manuscript,screen]{acmart}

\usepackage{booktabs}
\usepackage{graphicx,pstricks}
\usepackage{graphics}
\usepackage{lineno}
\usepackage{listings}
\usepackage{longtable}
\usepackage{makecell}
\usepackage{moreverb}
\usepackage{nicematrix}
\usepackage{seqsplit}
\usepackage{subfigure}
\usepackage{longtable}

\AtBeginDocument{%
  }

\setcopyright{none}
\copyrightyear{}
\acmYear{2025}
\acmDOI{XXXXXXX.XXXXXXX}
\acmConference[CHI '26]{ACM CHI Conference on Human Factors in Computing Systems}{April 13-17, 2025}{Barcelona, ES}
\acmISBN{978-1-4503-XXXX-X/2018/06}

\begin{document}

\title{Towards Better Health Conversations: The Benefits of Context-seeking}


\author{Rory Sayres}
\authornote{Both authors contributed equally.}
\authornote{Corresponding authors: sayres@google.com, mikeshake@google.com}
\author{Yuexing Hao}
\authornotemark[1]
\author{Abbi Ward}
\author{Amy Wang}
\author{Beverly Freeman}
\author{Serena Zhan}
\author{Diego Ardila}
\author{Jimmy Li}
\author{I-Ching Lee}
\author{Anna Iurchenko}
\author{Siyi Kou}
\author{Kartikeya Badola}
\author{Jimmy Hu}
\author{Bhawesh Kumar}
\author{Keith Johnson}
\author{Supriya Vijay}
\author{Justin Krogue}
\author{Avinatan Hassidim}
\author{Yossi Matias}
\author{Dale R. Webster}
\author{Sunny Virmani}
\author{Yun Liu}
\author{Quang Duong}
\author{Mike Schaekermann}
\authornotemark[2]
\affiliation{
  \institution{Google Research}
  \city{Mountain View}
  \country{Mountain View, USA}
}

\renewcommand{\shortauthors}{Sayres et al.}

\begin{abstract}
    Navigating health questions can be daunting in the modern information landscape. Large language models (LLMs) may provide tailored, accessible information, but also risk being inaccurate, biased or misleading. We present insights from 4 mixed-methods studies (total N=163), examining how people interact with LLMs for their own health questions. Qualitative studies revealed the importance of context-seeking in conversational AIs to elicit specific details a person may not volunteer or know to share. Context-seeking by LLMs was valued by participants, even if it meant deferring an answer for several turns. Incorporating these insights, we developed a “Wayfinding AI” to proactively solicit context. In a randomized, blinded study, participants rated the Wayfinding AI as more helpful, relevant, and tailored to their concerns compared to a baseline AI. These results demonstrate the strong impact of proactive context-seeking on conversational dynamics, and suggest design patterns for conversational AI to help navigate health topics.
\end{abstract}



\maketitle

\section{Introduction}

Modern information technology offers many resources for laypeople to understand their health concerns; but also poses potential pitfalls. People commonly use Internet search for their health and medical questions, with a recent survey reporting that 58.5\% of US-based adults used the Internet for this purpose \cite{wang2023health}. But questions persist about how people may interpret low-quality or less-relevant results and how it may influence population health behavior \cite{babic2021direct}. Meanwhile, large language models (LLMs) have been demonstrated to encode medical knowledge, and can perform well on a range of medical tasks \cite{singhal2023large, singhal2025toward, goodman2023accuracy}. A growing ecosystem of LLM-based tools has the potential to help provide more tailored or accessible information, and anecdotal reports have illustrated cases where LLMs surfaced crucial information for health conditions that were not captured by existing medical care \cite{HofmannGilmer2023Partners}. But this technology also has raised concerns about information accuracy and potential bias \cite{bean2025clinical, babic2021direct, hao2025large}. 

Addressing certain health information needs can require access to specific context that users may not proactively include in health searches. For instance, for cause-of-symptoms intent (henceforth COSI) needs, where a person wants to understand what is causing a set of symptoms, laypeople may not know what to proactively share to help ascertain the possible causes. Within clinical contexts, healthcare professionals are trained in history taking, and tested by evaluations such as the objective structured clinical examination \cite{engel1973interviewing, flugelman2021history}.

Recent research has developed LLMs for augmenting clinicians in tasks like reasoning about possible diagnoses, including in obtaining relevant context for differential diagnoses \cite{tu2025towards, mcduff2025towards}. However, there is little understanding of how LLMs can optimally support laypeople in their online health information seeking (OHIS) journeys in non-clinical settings. Because laypeople lack clinical training, the types of questions posed to the LLM, decisions being made, and the appropriate behavior of the LLMs may differ substantially. 

Our work examines how people interact conversationally with LLMs for their own health questions. We conducted 3 interview-based qualitative studies, and 1 quantitative study, to understand how participants engage with context-seeking in conversations about their own real-world health questions.

We make four key contributions:

\begin{enumerate}
    \item We provide qualitative evidence that proactive context-seeking by LLMs is valued by laypeople for their health questions, even if it delays an answer by several conversation turns.  
    \item We describe a \textbf{Wayfinding Artificial Intelligence (Wayfinding AI)} which was iteratively designed based on qualitative feedback to engage in effective context-seeking, to help users obtain the most relevant information for their concern.
    \item We compare user perceptions of the Wayfinding AI to the baseline LLM from which it was trained; in a quantitative study (N=130) we show that the Wayfinding AI is significantly preferred over the baseline AI on helpfulness, relevance of questions asked, tailoring, and efficiency of answering health questions.
    \item We discuss design considerations for how LLMs should behave in conversational contexts, to help users better provide relevant context and to comprehend and trust the outputs.
\end{enumerate}

\section{Related Work}

\subsection{Online Health Information Seeking}

Online health information seeking (OHIS) is a well-studied phenomenon in the human-computer interaction (HCI) literature. Prior work has explored OHIS in the context of different platforms including search engines and social media \cite{choudhyry2014seeking,yun2025framing}, specific medical domains like mental health \cite{milton2024seeking}, reproductive health \cite{dewan2024teen}, rare disorders \cite{chan2025odyssey} or COVID-19 \cite{di2024online}, as well as demographic factors like age, gender and ethnicity \cite{becker2004study,liao2014age,augustaitis2021transgender,harrington2022blackadults,harrington2023trust,dewan2024teen}. Research has found that the framing of information, algorithmic transparency, and overall system design can influence important aspects of user perception including trust, comfort and anxiety \cite{wu2025trust,yun2025framing,harrington2023trust}. Individuals' pursuit of health information is often driven by personal, emotional, and motivational factors and has been shown to take place within information ecosystems \cite{milton2024seeking}, rather than simply constituting a detached quest for facts \cite{jia2021online, lambert2007health,oberoi2016understanding}. In recent years, LLMs have become another popular starting point for OHIS journeys \cite{hao_advancing_2024}, and research has started probing for laypeople's attitudes towards LLMs versus other methods (like search engines) for their health queries \cite{mendel2025laypeople}.
Our work contributes to understanding how LLMs can optimally support laypeople in their OHIS journeys through proactive context-seeking behavior.

\subsection{Expert-layperson conversations}

When human experts in different domains converse with laypeople about their area of expertise, they adopt different strategies based on the layperson's questions and phrasing \cite{freeman2025zoom}: They will "zoom in" to gain context about specific questions when it is not initially provided, "zoom out" to clarify overall goals and explore other possible approaches; and "reframe" when question intent is ambiguous or relies on false assumptions.

In-person conversations, such as patient-provider discussions, can rely on a range of nonverbal cues to gauge understanding of content and level of concern \cite{bromme2005barriers}. As more conversations are done using text or other remote interfaces, experts have had to adapt conversational strategies to better assess these facets of communication \cite{li2023constraints, li2025comparative}. However, some of these strategies make conversation harder for laypeople, such as the use of "double-barrel" questions that are harder to answer concisely \cite{li2023constraints}.

\subsection{Clinician-AI conversations}

Most clinical decision support AI has focused on reading of medical images \cite{sutton2020overview, esteva2019guide}. Use patterns include risks of over-reliance and under-reliance \cite{tschandl2020human, gaube2021ai, parasuraman1997humans, detjen2025trusted}. Earlier applications tended to involve AI in narrowly-defined tasks, such as reading specific image modalities (X-rays, mammograms, retina images) for particular diseases \cite{mckinney2020international, sayres2019using}. 
More recent work has examined how conversation may be used to support clinicians across a wider range of clinical domains, including differential diagnosis based on understanding complex information and history taking \cite{mcduff2025towards, tu2025towards, saab2025advancing}.

\subsection{Layperson decision support with AI tools}

Consumer-oriented, AI-powered apps are available in some countries for uses such as detecting common types of skin cancers \cite{smak2023artificial}. Use of an app of this kind was associated with an increase in claims for skin malignancies in the health care system, suggesting greater detection (and potentially over-detection) of cancers \cite{smak2023artificial}. AI tools to support understanding skin concerns based off smartphone images were helpful for laypeople \cite{krogue2024searching, sayres2025navigating}, could increase their ability to name a condition and may support patient-provider interactions \cite{sayres2025navigating}; but people tended to anchor on the example images shown by the app, suggesting the app design may strongly impact app benefits.

Many people use Internet search to find information about health topics online \cite{wang2020human}. However, the risks and benefits of existing web search patterns are not well characterized. A study of Internet search for health information, diagnosis and triage on retrospective vignettes found that any benefit of Internet search was small, with a small fraction of respondents changing their diagnosis or triage decisions after Internet search \cite{levine2021assessment}.

\subsection{Understanding conversational patterns with health chatbots}

As more people have begun using conversational AI chatbots, they have started asking these tools their health questions \cite{ayre2025use, shahsavar2023role, fan2021utilization}. Individual stories suggest these interactions may be highly valuable in some cases \cite{HofmannGilmer2023Partners}; but may pose a range of risks \cite{Reiley2025,de2024health, eichenberger2025case}. 
Several studies suggest broad interest in health-related AI chatbots \cite{shahsavar2023role, chang2022would, nadarzynski2019acceptability, ayre2025use}. Survey studies found a large proportion of respondents who either intended (78\%, \cite{shahsavar2023role}) or recently had (9.9\%, \cite{ayre2025use}) used ChatGPT for their health questions, with a highly significant effect of age (younger respondents using it more often).
There has been a focus within the HCI literature on how laypeople have conversations with AI chatbots on their own health concerns. A study on chatbot-based symptom checkers identified a tension between a desires for efficient information access, and one for empathetic communication \cite{you2023beyond}. Several studies used a "Wizard-of-Oz" design, with human experts powering AI responses to understand layperson conversational behavior \cite{li2023constraints, li2024beyond, li2024conversational, li2025comparative}. People chatting with clinical experts (dermatologists) disguised as a chatbot will exhibit different patterns of conversational turns compared to discussions with LLMs \cite{li2024conversational}, with more messages overall, as well as more expressions of appreciation and question answering. Likewise, in a clinical pre-consultation setting, expressions of empathy and follow-up questions impacted users' perception of a chatbot's thoroughness and sincerity \cite{li2025comparative}. These suggest that the default conversational patterns of current LLMs may not sufficiently mirror the behaviors of human experts.

\subsection{Need for understanding real-world usage patterns}

Despite growing numbers of people conversing with AI chatbots, little is known about what successful interaction looks like, especially on health topics. Controlled studies on a set of hand-selected medical scenarios showed that, when LLMs were given scenarios directly, they achieved high accuracy; but when laypeople were assigned to act out the scenarios with the same LLMs, performance at assessing conditions plummeted by more than half, to the same levels as without access to LLMs \cite{bean2025clinical}. Similar results showed that LLMs would often "get lost" in multi-turn settings, exhibiting much lower performance when provided context across turns compared to up front \cite{laban2025llms}. The large gap in performance derived from difficulty of users in providing necessary context around each scenario. Analyses of LLM architectures suggests that human conversations may pose a particular challenge compared to other domains like software and mathematical text, due to the long-range attention across many conversation turns required \cite{jawale2024human}. Recent research has focused on developing frameworks to quantify aspects of conversation, such as the quality of clarifying questions \cite{li2024mediq, li2025alfaaligningllmsask, johri2024craft}.

These prior results point to conversation dynamics, and the need for context-seeking in particular, as key design factors influencing successful interaction with an AI. To our knowledge, there has not yet been a study of how laypeople use functional AIs (rather than Wizard-of-Oz simulations) on their own health problems. Do people experience the same challenges in providing necessary context to get relevant answers on their own concerns as on simulated scenarios \cite{bean2025clinical}? And do their conversational patterns with AIs mirror those of humans presented as AIs \cite{li2024conversational, li2025comparative}? And are these prior insights actionable -- does designing an AI to behave more like human experts lead to better conversations? This study sought to focus on understanding these real-life interaction patterns.

\section{Methods}

An overview of the 4 studies in this paper is provided in Table \ref{tab:studies_summary}. The qualitative studies (Studies 1-3) and quantitative study (Study 4) are described in further detail below.

\begin{table}[ht]
    \centering
    \begin{tabular}{m{1cm}|m{7.5cm}|m{3.1cm}|c}
    \toprule[2pt]
    \textbf{Study} & \textbf{Key Research Questions} & \textbf{Methodology} & \textbf{\makecell[l]{Num of \\Participants}}\\
    \midrule[2pt]
    Study 1 & \begin{enumerate}
        \item How do people view AIs' context-seeking behaviors?
        \item Is context-seeking valued even if it means deferring an answer for several conversation turns?
    \end{enumerate} & Interviews, remote usability session & 10 \\
    \midrule[1.5pt]
    Study 2 & \begin{enumerate}
        \item How do people view a Wayfinding AI compared to a Baseline AI when using it for their health questions?
        \item How is context-seeking perceived on models that tend to give longer responses by default?
    \end{enumerate} & Interviews, remote usability session & 11 \\
    \midrule[1.5pt]
    Study 3  & \begin{enumerate}
        \item Does a user interface that explicitly separates answers from context seeking increase engagement with the AI?
        \item What are user perspectives on different UI affordances for context seeking?
    \end{enumerate} & Interviews, remote usability session & 12 \\
    \midrule[1.5pt]
    Study 4 & \begin{enumerate}
        \item Is a Wayfinding AI preferred over a Baseline AI for conversations on health questions?
        \item How do conversation dynamics differ between Wayfinding AI and Baseline AIs?
    \end{enumerate} & Randomized blinded unmoderated survey study with remote chatbot use & 130 \\    
    \bottomrule[2pt]
    \end{tabular}

    \caption{Summary of the studies described in this paper.}
    \label{tab:studies_summary}
\end{table}

\subsection{Ethics \& Positionality Statement}

This study was meant to focus on user experience, and participants were instructed not to consider information from the research tool as advice or to act on it. The protocol was reviewed and deemed exempt from Institutional Review Board (IRB) oversight by the Advarra IRB. The study was conducted in adherence to our organization's ethical, legal, and privacy standards for human subjects research. 

\subsection{AI prototypes}

This work involved participants interacting with a series of AI prototypes to understand their own health questions. 

\textbf{Wayfinding AI}: The primary goal was to develop and test a Wayfinding AI to proactively seek out context of user’s health queries, in order to provide a maximally relevant and accurate answer. The Wayfinding AI was developed using a combination of prompt tuning and reinforcement learning off of the baseline AI. Over the course of the 4 studies described in this paper (3 qualitative, 1 quantitative), we iteratively refined our Wayfinding AI to better address feedback from each qualitative study. For the final quantitative study, the Wayfinding AI configuration was frozen prior to commencing the study. We describe the full set of AIs across studies in supplemental Table \ref{tab:agents_summary}.

\textbf{Baseline AI}: In Studies 2 and 4, we compared the Wayfinding AI to a Baseline AI. Our baseline AI utilized Gemini 2.5 Flash \cite{comanici2025gemini} off-the-shelf, without any additional tuning or system prompts specifically designed to engage in context-seeking. This AI served as a comparison baseline in our blinded, randomized Study 4.

AIs were presented to users as chatbots in a custom interface (Figure \ref{fig:AI_screenshots}) which presented alternating user prompts and system responses. For one version of the interface, used in qualitative Study 3 and the quantitative Study 4, the interface used a 2-column design that surfaced the best answer using available information so far in a right-hand column, leaving only clarifying questions in the left (main) column with the user-AI conversation, in order to provide a clear affordance for context-seeking.

\subsection{Qualitative studies (Studies 1-3)}

Three rounds of qualitative studies took place from February to June 2025. We recruited a total of 33 US-based participants for 3 interview-based qualitative studies (10 in Study 1, 11 in Study 2, and 12 in Study 3). Participants ranged in age from 25-75; 17 female / 17 male. Participant demographics included 19 Caucasians, 6 Asian or Pacific Islanders, 4 Hispanics, and 4 Black or African Americans. The full list of participant information and their health questions are presented in Table \ref{tab:R1_participants}. 

All participants had a recent health-related query they were willing to discuss, and were open to interacting with one or more AIs for the study. Sessions lasted 1 hour each and had the following structure: First, participants discussed their question, what information sources they had used so far to understand it, and their current status of understanding the question. Second, participants engaged in a remote usability session while they interacted with an AI to better understand the topic of their question. Researchers probed on why participants provided the prompts they did; what their expectations were for AI responses; what information they hoped to obtain; and whether the AI’s behavior matched their expectations. Third, participants used a second, alternate AI. The order of AIs tested by each participant was randomized across participants for each study. Finally, participants compared the two AIs and provided overall impressions and thoughts about their experience and what they learned.

\begin{table}
    \begin{tabular}{lllllllll}
    \toprule[2pt]
    \textbf{ID} & \multicolumn{1}{l}{\textbf{Gender}} & \multicolumn{1}{l}{\textbf{Age}} & \multicolumn{1}{l}{\textbf{Race / Ethnicity}} & \multicolumn{1}{l}{\textbf{Health Question(s) Summary}} \\ \midrule[2pt]
    
    P1 & \multicolumn{1}{l}{Female} & \multicolumn{1}{l}{47} & \multicolumn{1}{l}{Hispanic} & \multicolumn{1}{l}{Consistent and unexplained weight gain} \\
    P2 & \multicolumn{1}{l}{Female} & \multicolumn{1}{l}{26}  & \multicolumn{1}{l}{Asian/Pac Is} & \multicolumn{1}{l}{Respiratory symptoms after travel} \\
    P4 & \multicolumn{1}{l}{Male} & \multicolumn{1}{l}{50} & \multicolumn{1}{l}{Cauc} & \multicolumn{1}{l}{Tinnitus and surgery} \\
    P5 & \multicolumn{1}{l}{Male} & \multicolumn{1}{l}{62}  & \multicolumn{1}{l}{Asian/Pac Is} & \multicolumn{1}{l}{Rash diagnosis and treatment} \\
    P6 & \multicolumn{1}{l}{Female} & \multicolumn{1}{l}{59}& \multicolumn{1}{l}{Cauc} & \multicolumn{1}{l}{Cataract surgery} \\
    P7 & \multicolumn{1}{l}{Male} & \multicolumn{1}{l}{25} & \multicolumn{1}{l}{Cauc} & \multicolumn{1}{l}{Torso itchy when warm} \\
    P8 & \multicolumn{1}{l}{Male} & \multicolumn{1}{l}{46} & \multicolumn{1}{l}{Cauc} & \multicolumn{1}{l}{Medication's long-term side effects} \\
    P9 & \multicolumn{1}{l}{Female} & \multicolumn{1}{l}{36} & \multicolumn{1}{l}{AA} & \multicolumn{1}{l}{Long-lasting migraines and the causes} \\
    P10 & \multicolumn{1}{l}{Male} & \multicolumn{1}{l}{40} & \multicolumn{1}{l}{Cauc} & \multicolumn{1}{l}{Essential thrombocytosis} \\
    \bottomrule[2pt]
    P11 & \multicolumn{1}{l}{Male} & \multicolumn{1}{l}{47} & \multicolumn{1}{l}{Cauc} & \multicolumn{1}{l}{Reverse or lessen tinnitus} \\
    P12 & \multicolumn{1}{l}{Male} & \multicolumn{1}{l}{31} & \multicolumn{1}{l}{Asian / Pac Is} & \multicolumn{1}{l}{Stress-induced dandruff} \\
    P13 & \multicolumn{1}{l}{Female} & \multicolumn{1}{l}{68} & \multicolumn{1}{l}{Asian / Pac Is} & \multicolumn{1}{l}{Blood pressure and sleep issues} \\
    P14 & \multicolumn{1}{l}{Male} & \multicolumn{1}{l}{65} & \multicolumn{1}{l}{Cauc} & \multicolumn{1}{l}{Natural blood pressure supplement} \\
    P15 & \multicolumn{1}{l}{Male} & \multicolumn{1}{l}{54} & \multicolumn{1}{l}{Asian / Pac Is} & \multicolumn{1}{l}{Rib pain} \\
    P16 & \multicolumn{1}{l}{Male} & \multicolumn{1}{l}{57} & \multicolumn{1}{l}{Asian / Pac Is} & \multicolumn{1}{l}{Daily headaches} \\
    P17 & \multicolumn{1}{l}{Female} & \multicolumn{1}{l}{33} & \multicolumn{1}{l}{Cauc} & \multicolumn{1}{l}{Child's fussiness and infertility concerns} \\
    P18 & \multicolumn{1}{l}{Female} & \multicolumn{1}{l}{41} & \multicolumn{1}{l}{Cauc} & \multicolumn{1}{l}{Perimenopause symptoms and Hormone replacement therapy} \\
    P19 & \multicolumn{1}{l}{Female} & \multicolumn{1}{l}{56} & \multicolumn{1}{l}{Hispanic} & \multicolumn{1}{l}{Osteoporosis} \\
    P20 & \multicolumn{1}{l}{Male} & \multicolumn{1}{l}{29} & \multicolumn{1}{l}{AA} & \multicolumn{1}{l}{Swelling causes} \\
    P21 & \multicolumn{1}{l}{Female} & \multicolumn{1}{l}{49} & \multicolumn{1}{l}{Cauc} & \multicolumn{1}{l}{Pain and sensitivity in leg} \\
    \bottomrule[2pt]
    P22 & \multicolumn{1}{l}{Male} & \multicolumn{1}{l}{37} & \multicolumn{1}{l}{Cauc} & \multicolumn{1}{l}{Eosinophil Esophagitis: Swallowing issues} \\
    P23 & \multicolumn{1}{l}{Female} & \multicolumn{1}{l}{25} & \multicolumn{1}{l}{Hispanic} & \multicolumn{1}{l}{Frequent, sharp abdominal pain} \\
    P24 & \multicolumn{1}{l}{Male} & \multicolumn{1}{l}{38} & \multicolumn{1}{l}{Hispanic} & \multicolumn{1}{l}{Knee issues: clicking, locking, pain} \\
    P25 & \multicolumn{1}{l}{Male} & \multicolumn{1}{l}{61} & \multicolumn{1}{l}{AA} & \multicolumn{1}{l}{Suspected kidney stones and prevention tips} \\
    P26 & \multicolumn{1}{l}{Female} & \multicolumn{1}{l}{25} & \multicolumn{1}{l}{Cauc} & \multicolumn{1}{l}{Vertigo} \\
    P27 & \multicolumn{1}{l}{Male} & \multicolumn{1}{l}{45} & \multicolumn{1}{l}{Cauc} & \multicolumn{1}{l}{Lower back pain cause} \\
    P28 & \multicolumn{1}{l}{Female} & \multicolumn{1}{l}{60} & \multicolumn{1}{l}{Cauc} & \multicolumn{1}{l}{Knee pain, buckling, and exercise} \\
    P29 & \multicolumn{1}{l}{Female} & \multicolumn{1}{l}{53} & \multicolumn{1}{l}{Cauc} & \multicolumn{1}{l}{Cyst symptoms reoccurring after surgery} \\
    P30 & \multicolumn{1}{l}{Male} & \multicolumn{1}{l}{48} & \multicolumn{1}{l}{Cauc} & \multicolumn{1}{l}{Seed oils and gout flare-ups} \\
    P31 & \multicolumn{1}{l}{Female} & \multicolumn{1}{l}{75} & \multicolumn{1}{l}{Cauc} & \multicolumn{1}{l}{Acid reflux: symptoms and remedies} \\
    P32 & \multicolumn{1}{l}{Female} & \multicolumn{1}{l}{33} & \multicolumn{1}{l}{Cauc} & \multicolumn{1}{l}{Ovarian cysts} \\
    P33 & \multicolumn{1}{l}{Male} & \multicolumn{1}{l}{50} & \multicolumn{1}{l}{Cauc} & \multicolumn{1}{l}{Raynaud's syndrome and discolored feet} \\
    \bottomrule[2pt]

\end{tabular}    
\vspace{0.2cm}
\caption{Study 1-3 Participant Demographic Information.}
\label{tab:R1_participants}
\end{table}

\begin{figure*}[!ht]
    \centering
    \includegraphics[width=0.6\textwidth]{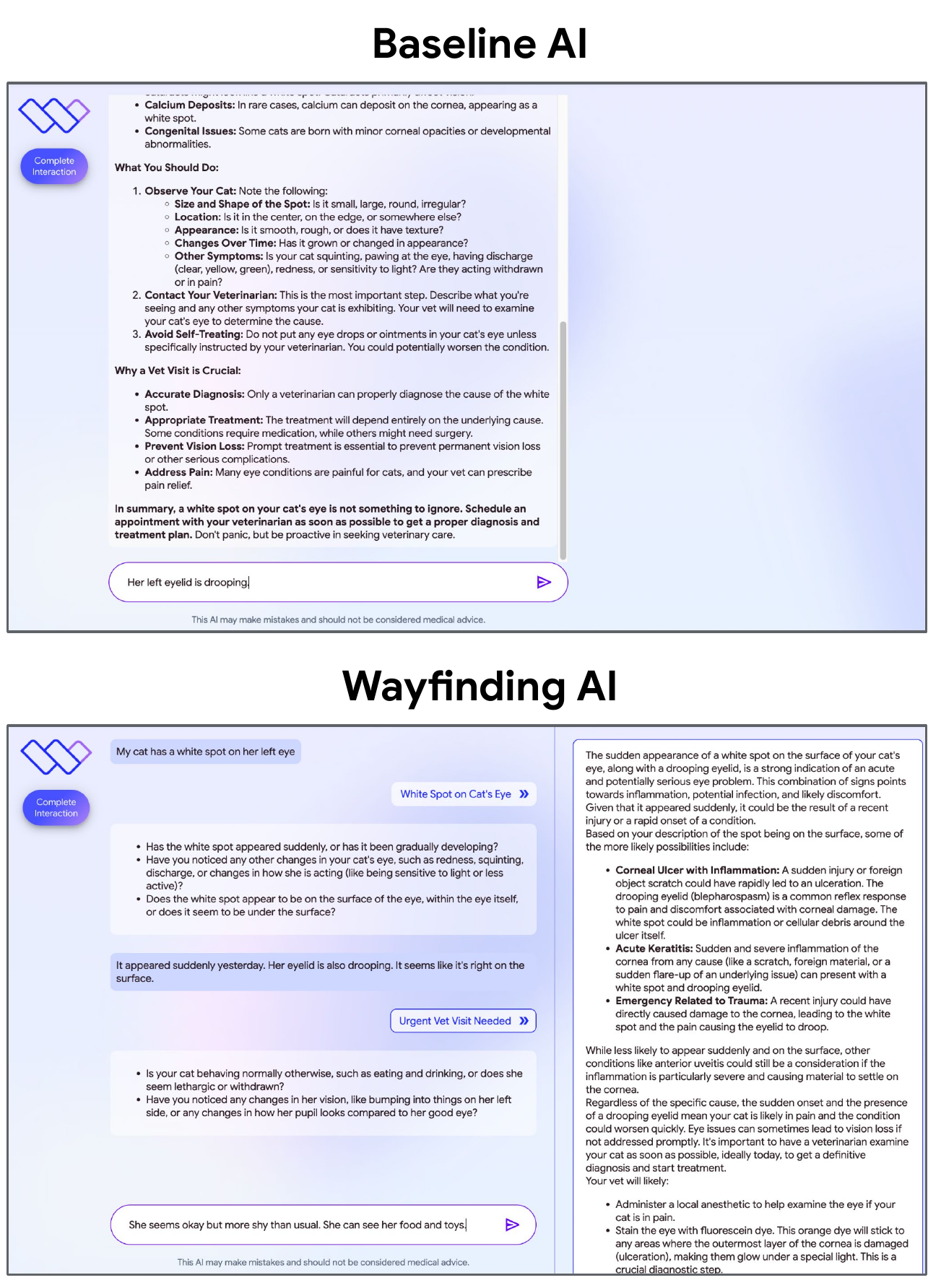}
    \caption{Screenshots of the two AI chatbot designs used in Studies 3 and 4. Top shows a baseline, 1-column AI design. Bottom shows a 2-column design in which the current answer is shown in the right column, while clarifying questions are displayed underneath user prompts on the left.}
    \label{fig:AI_screenshots}
    \Description{This figure provides illustrative screenshots of the research-only AI AI chatbots used in the study. The top shows a Baseline AI, with a conversation between a user and an AI AI about a hypothetical abdominal pain issue. User turns are shown against a darker purple background, while AI responses are shown against a lighter blue background. The example AI response includes both text providing an answer, and some clarifying questions below it, all in one text block. There is a prompt area at screen bottom with placer text to "Type Anything here", where users can add additional prompts. There is a button to "Complete Interaction" at the screen left, and a disclaimer text at bottom stating "This AI may make mistakes and should not be considered medical advice." The bottom shows a Wayfinding AI, with a discussion on the same topic. The Wayfinding AI has the same overall appearance and layout as the Baseline AI, but there is a separate right-hand panel in which the bulk of AI responses are shown. The answer portion of the AI response is shown in the right panel, while the clarifying questions are beneath the user's prompt on the left.}
\end{figure*}

The AIs studied varied between the three studies, based on the key research questions (see Table \ref{tab:agents_summary}). Study 1 compared a Wayfinding AI that provided an immediate answer to a question to a modified version with a system prompt to ask at least 4 but no more than 10 clarifying questions before providing an answer. The prototypes in Study 1 were specific to cause-of-symptom-intent (COSI) questions, so for this study we recruited participants who specifically reported having a question involving understanding symptoms. Study 2 compared a tuned Wayfinding AI (that was designed to handle both COSI and non-COSI questions) to the base AI (without tuning). For this study, we recruited participants with a wider range of questions, including both cause-of-symptom as well as other intents (e.g., treatment or management for a known condition). Study 3 used a further refined Wayfinding AI, and compared two alternative UI arrangements, in which answers to questions were either placed in-line and above context-seeking questions (1-column layout) or in a separate side panel (2-column layout). For Study 3, we again focused on COSI questions, in order to focus our feedback on how the UI layouts facilitated context seeking for this common use case. The different UI layouts are illustrated in Figure \ref{fig:AI_screenshots}.

\subsection{Quantitative study (Study 4)}

\begin{figure*}[!htp]
    \centering
    \includegraphics[width=\textwidth]{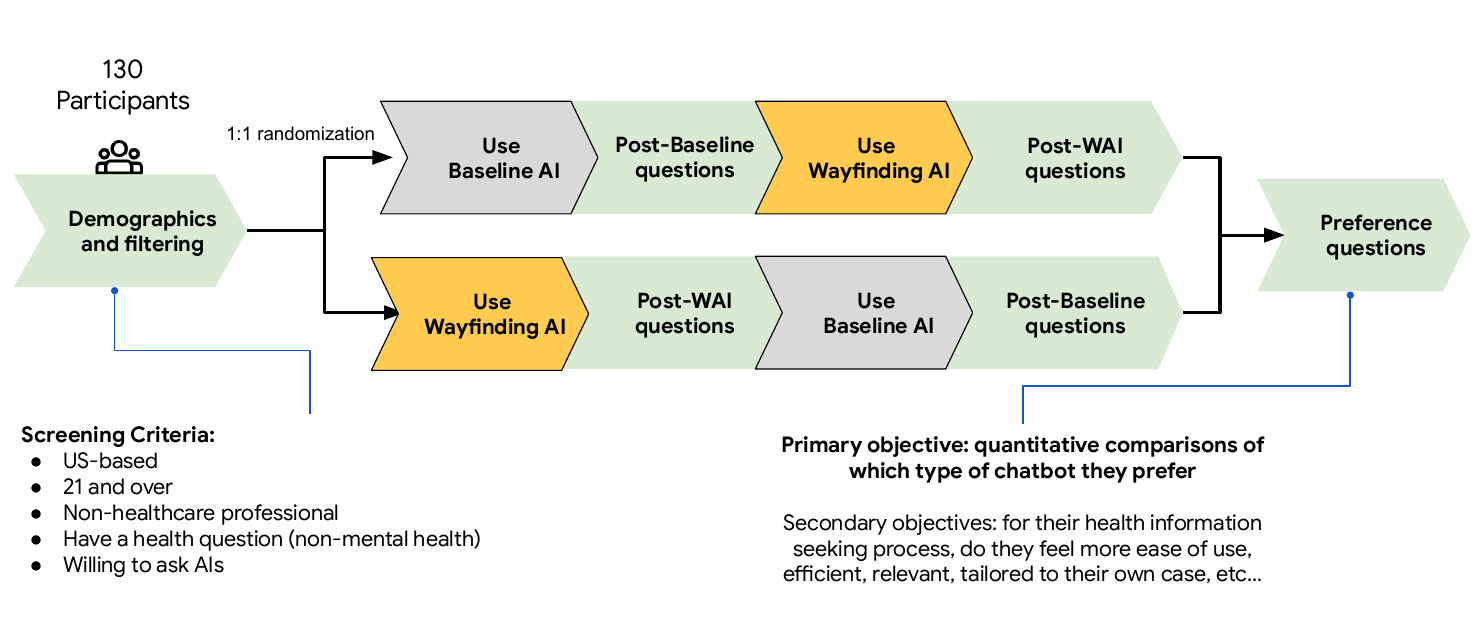}
    \caption{Quantitative study design (Study 4). All participants used both AIs used in the study (Wayfinding AI and Baseline AI) to better understand their own health-related questions; but were randomly assigned in the order with which to use them. After using each AI, participants answered questions about their satisfaction with the experience; at the end of the survey, subjects explicitly compared the two AIs and expressed which they would prefer along a range of axes.}
    \label{fig:study_design}
    \Description{This figure is a flowchart illustrating the design of the quantitative study. 130 participants provided demographic information and verified that they are eligible to participate in the study. Eligibility criteria are marked in the flowchart: US-based; 21 and over; Non-healthcare professional; Have a health question (non-mental health); Willing to ask AIs. Participants were then randomized: Half used the Baseline AI, answered post-Baseline questions, then used the Wayfinding AI and answered post-Wayfinding AI questions. The other half used the Wayfinding AI, then the Baseline AI. After using both AIs, they answered preference questions about which AI they preferred for different axes of use. The flowchart is annotated to note "Primary objective: quantitative comparisons of which type of chatbot they prefer; Secondary objectives: for their health information seeking process, do they feel more ease of use, efficient, relevant, tailored to their own case, etc."}
\end{figure*}

We assessed user satisfaction and preference between Wayfinding AI and Baseline AI in a quantitative survey-based study (Figure \ref{fig:study_design}). We recruited US-based participants via a third party platform (Qualtrics). All participants were 21 years and older, were not health care professionals, and had a health-related question for which they were willing to interact with an AI. Mental-health-related queries were excluded from the study scope. 

After answering standard demographic questions, participants were instructed to login to each AI using anonymous accounts set up specifically for this study; to have a conversation spending at least 3 minutes on their question; and then to resume the survey. After interacting with each AI, participants answered questions about their satisfaction with the experience along 6 dimensions: Helpfulness, relevance of questions asked, tailoring to their situation, goal understanding, ease of use, and efficiency of getting useful information; and could provide open feedback about what they learned, as well as the option to upload their conversation with the AI. Sharing the conversation was not required to complete the survey. At the end of the study, participants were prompted to explicitly compare the two AIs and indicate which they would prefer in terms of each of the six dimensions above, and answered an additional question: "For a future topic, would you prefer the first or the second AI?". The order of AI exposure (Baseline AI first vs Wayfinding AI first) was randomized across participants. 

\subsection{Analysis}

\subsubsection{Qualitative study analysis}

All qualitative interview sessions (Studies 1-3) were recorded. Recordings were transcribed using a text-to-speech medical transcription model. Transcripts were analyzed using reflexive thematic analysis \cite{clarke2014thematic}. In addition, transcripts of conversations with each AI were analyzed to understand turn structure. Conversations were broken into a set of alternating user and AI turns; user turns were manually coded to determine whether responses directly answered clarifying questions asked by each AI.

\subsubsection{Quantitative study analysis}

Quantitative data analyses were performed using Python 3.0 and the SciPy package. Confidence intervals on measurements were computed using bootstrapping with 1,000 iterations. Paired comparisons on survey responses between AIs used the Wilcoxon signed-rank test for ordinal responses, and paired two-sided T tests for interval responses. Significance on comparison tests used Bonferroni correction for multiple comparisons across different aspects of user satisfaction and/or preference.

Participants in the quantitative study could optionally upload their conversations with each AI along with their survey responses. Conversations were screened for personally-identifiable information by Qualtrics before sharing with researchers. A total of 119 conversations for the Wayfinding AI and 112 for the Baseline AI were shared. Each conversation was broken into turns where user prompts and AI responses were analyzed. We prompted Gemini 2.5 Pro \cite{comanici2025gemini} to classify user prompts and AI responses in multiple ways for analysis. Initial user prompts were classified as reflecting cause-of-symptom intent (COSI; e.g. "a cough that won't go away") or non-cause-of-symptom intent (NCOSI; e.g. "I am curious to know about keto"), to understand how specific the benefits of the Wayfinding AI might be to COSI. All user prompts were also classified in one of 11 turn types, in order to understand conversational dynamics. The classification scheme was derived by internal review and discussion among researchers. Finally, AI responses were classified as containing clarifying questions (e.g. "Have you ever had chickenpox or the shingles vaccine previously?"), open-ended questions ("Could you tell me a little about what you hope to achieve with your workout plan?") or no questions. All prompts used to classify conversational turns are included in the Appendix.

\section{Results}

\subsection{Study 1: Perceived value of context-seeking}

Participants in Study 1 consistently found value in an AI that delayed answers for 4 or more turns  to collect context around questions. 8 of the 10 participants stated that they preferred to use a prototype that only asked clarifying questions for 4 turns before providing an answer (AI 1b) over a prototype that provided immediate answers before context-seeking (AI 1a). Participants called out the value of context-seeking in establishing a conversational interaction that felt more human (P2: “I feel good. I feel like I’m having real conversations with someone, as opposed to me trying to figure out the level of information that I need to give ... that’s one of my main issues, sometimes.” P7: “This one feels more like the way it would work if you talk to a doctor. Conversational ... it does make make me feel a little more confident that it wants to know more before jumping right into an answer.”) Conversations took between 2 to 6 turns for AI 1a, and between 4 and 16 turns for AI 1b. 9 of 10 participants directly answered one or more clarifying questions from 1a, and all 10 answered clarifying questions from 1b.

While participants found context-seeking useful for understanding possible matching conditions for symptoms, they also expressed other needs related to their health questions that the initial Wayfinding AI prototypes did not serve as well. Many participants wanted to understand management options for conditions; since the initial AIs in this study focused on cause-of-symptom intent, it took more prompting to get useful information about management. Therefore we sought to refine our Wayfinding AI to one that could more flexibly provide context-seeking for both COSI and non-COSI needs. Because non-COSI needs may inherently involve less zooming-in, we opted to focus on testing Wayfinding AIs that surfaced immediate answers in addition to context-seeking.

\subsection{Study 2: Tension between longer answers and context-seeking}

Study 2 involved 11 participants, who had both COSI and non-COSI questions. Participants found AI 2a (with context-seeking behavior) to be helpful for both use cases. Overall, 6 participants preferred 2a, 3 preferred 2b, and 1 had no preference. Participants who preferred AI 2b cited its context-seeking behavior, which helped them arrive at more relevant results. Participants preferring the baseline AI generally did so because they expressed that their needs were different and less likely to require “zooming in”, and they found the baseline AI responses to be comprehensive. These participants tended to have short conversations, with only 1 or 2 turns.

We noted an unexpected behavior in Study 2 compared to Study 1: only 1 of 10 participants in Study 2 directly answered clarifying questions asked by AI 2a (whereas all or nearly all participants did so for 1a and 1b). More often, participants would either pivot to related but distinct questions (e.g. P12: “What is the best diet for healthy scalp and hair?” $\rightarrow$ “Does lack of sleep lead to vitamin deficiencies?”); proactively zoom in to learn more details of a suggested topic (e.g. P18: “what are recommendations for treatment of perimenopause?” $\rightarrow$ “What are the risks and benefits of HRT in perimenopause?”) or correct the model when its questions seemed to have an inaccurate premise (e.g. P19:  “I'm a 56-year-old woman. Be more specific for what I need to do”).

When prompted, participants indicated that this was because they did not always notice the questions, or perceive them as essential for getting relevant answers. (P11: “They’re just questions for me to ask myself and to think about… It just ends with some rhetorical questions to consider.”) AI 2b tended to produce long answers (longer than 1a or 2a), and the clarifying questions were not as consistently placed, sometimes appearing at the end of an answer, but sometimes in the middle. Participants were more likely to report scanning these longer answers for relevant information, whereas participants in Study 1 more often read the whole answers. However, when probed about answer length, participants tended to report that they valued the longer and more comprehensive answers. (P12: “I do quite like this ... it looks like it’s trying more to not just give standard answers.”)

Due to the dual feedback of (1) desire for immediate, detailed answers and (2) challenges with consistent  context-seeking, we explored UI approaches to provide affordances for both.

\subsection{Study 3: Separating answers from context-seeking}

In Study 3, 12 participants used a refined Wayfinding AI that surfaced clarifying questions consistently at the end of answers, presented in two different user interface (UI) layouts. AI 3a placed answers directly above clarifying questions in the main column, while AI 3b placed them on the right-hand side in a two-column format. AI 3b had two variants, where the initial answer was either collapsed or expanded by default. The answer could be expanded by clicking on a clickable button (whose text provided a short summary of the response) in the main column. Participants would test one of the variants randomly, but be shown the second variant to get feedback on their preferences for expanded vs. collapsed behavior, as well as between 3a and 3b.

Participants were split on preference between 1-column and 2-column layouts. 6 preferred 3b, 4 preferred 3a, and 1 had no preference. Participants who preferred 3b cited how separating out clarifying questions encouraged a more naturalistic conversation (P27: “These are logical questions. They're on topic. And I could see how they would help the AI learn more about my problem to give me better responses.”) However, participants uniformly preferred the expanded variant, with some participants not initially recognizing that the answers could be expanded. Participants who preferred the 1-column view cited the simpler mental model associated with it (P31: “Right off, I like this version better than the side one ... You want to ask a question, then you want the answer down below. It’s just less confusing.”).

\subsection{Study 4: Wayfinding AI preferred over baseline AI}

Based on the qualitative studies, we developed AI 4a, a Wayfinding AI that provided answers to health questions as well as context-seeking in separate columns (with answers always visible in an expanded state). We compared this Wayfinding AI to a baseline AI (4b) in a quantitative study with US-based participants, who tested both models on their own health questions.

Of the 130 participants, 59\% were 50 years of age or younger and 44\% identified as female. The participant sample included various racial/ethnic groups (8\% Asian, 10\% Black or African American, 12\% Hispanic, Latino or Spanish Origin, 72\% White). Slightly more than half (61\%) of participants had earned a Bachelor's degree or a higher qualification. The majority (73\%) reported feeling extremely confident using mobile phone apps by themselves, and a smaller portion (61\%) reported the same level of confidence filling out medical forms by themselves. Slightly less than half (43\%) said they used AI chatbots at least several times a week, and 15\% reported never having used an AI chatbot before.  A detailed breakdown of participant characteristics is provided in Table \ref{tab:study_4_participants}.

\begin{figure*}[ht]
    \centering
    \includegraphics[width=300pt]{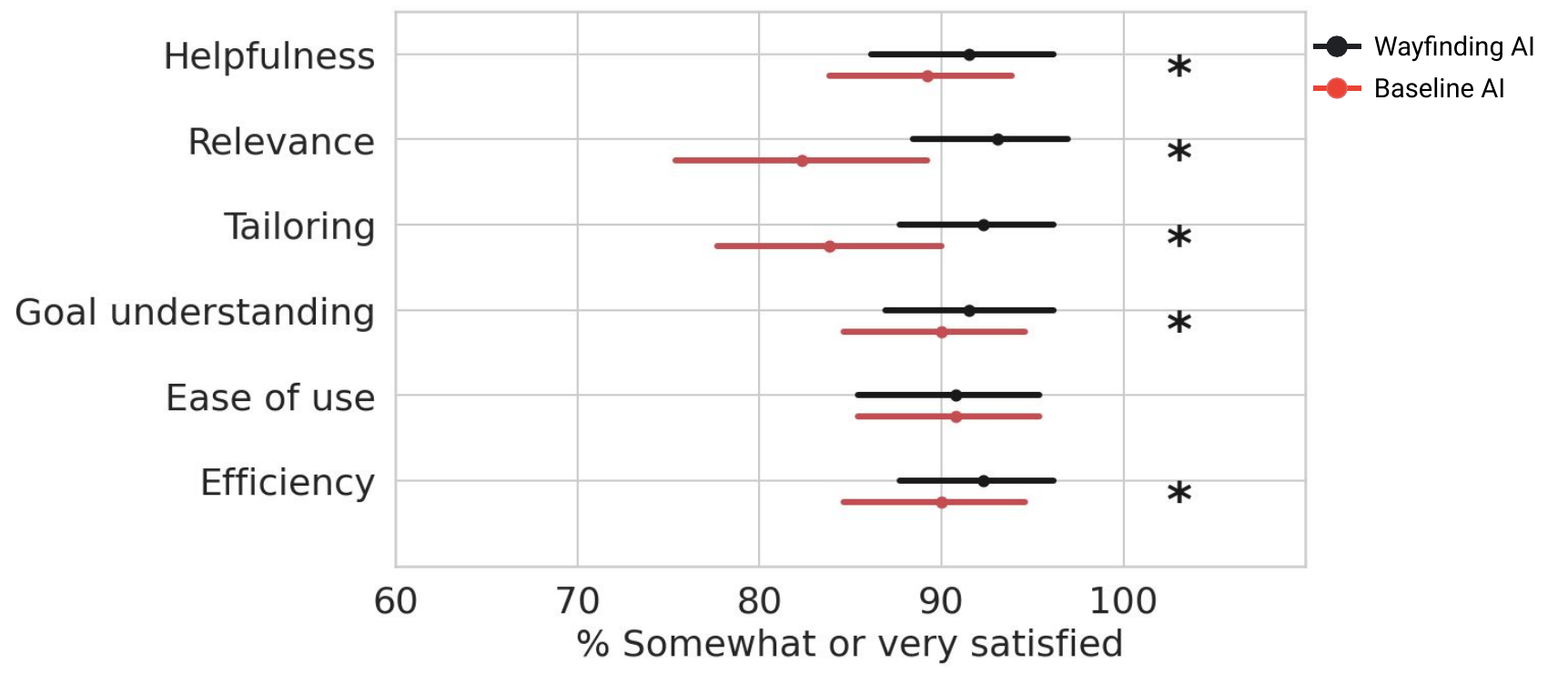}
    \caption{Summary of satisfaction measures for Wayfinding AI and Baseline AI. Values reflect the percent of respondents who report being somewhat or very satisfied with each agent. Error bars represent 95\% confidence intervals determined by bootstrap. Asterisks indicate significant difference by a Wilcoxon test, at p < 0.05, with Bonferroni correction for multiple comparisons.}
    \label{fig:satisfaction_summary}
    \Description{This chart shows satisfaction rates with each AI. Data points mark the mean rate of participants being very somewhat or very satisfied with the Wayfinding and Baseline AIs; error bars indicate 95\% confidence intervals by bootstrap. Asterisks indicate significant differences betwen the two AIs. Satisfaction is significantly higher for the Wayfinding AI than the Baseline AI along dimensions of Helpfulness, Relevance, Tailoring, Goal Understanding, and Efficiency, but not Ease of Use.}
\end{figure*}

When asked about their satisfaction with each app separately, participants reported higher satisfaction with the Wayfinding AI on the dimensions of helpfulness, relevance of questions asked, tailoring of responses to the participants’ situations, understanding of user goals, and efficiency of getting an answer (Figure \ref{fig:satisfaction_summary}).

\begin{figure*}[!ht]
    \centering
    \includegraphics[width=\textwidth]{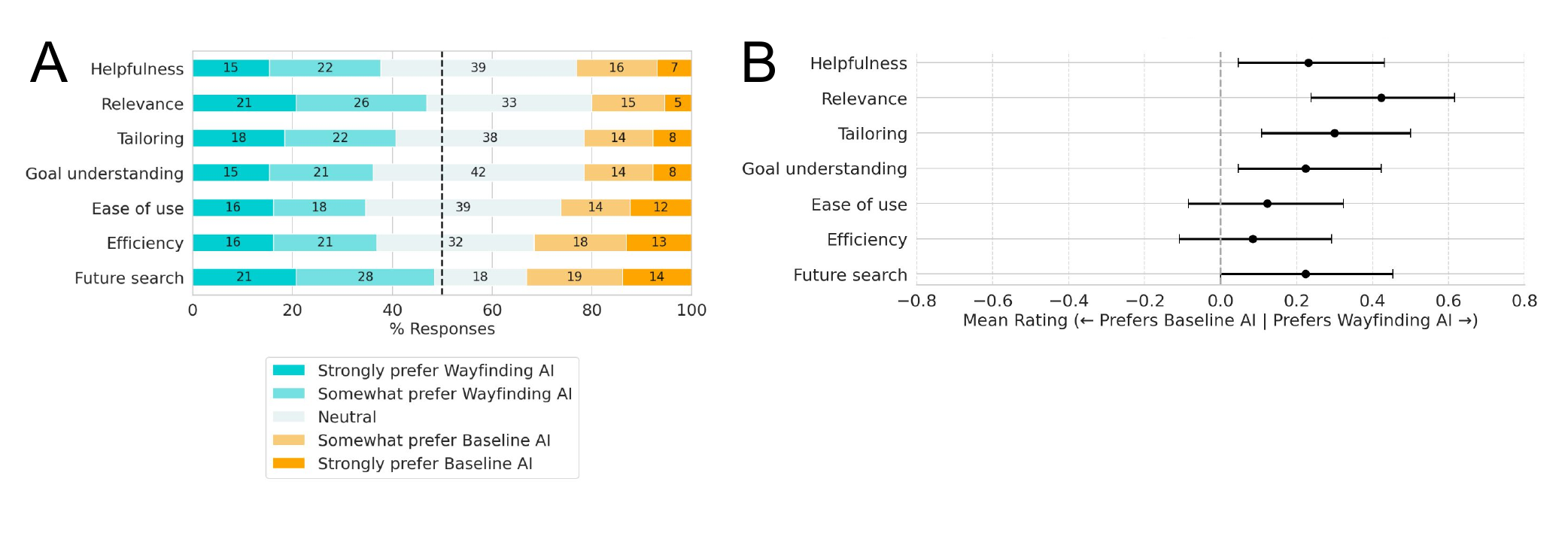}
    \caption{Summary of side-by-side comparison between Wayfinding AI and Baseline AI. \textbf{A}, distributions of survey responses for each preference question; \textbf{B}, Summary of preference distributions. Values reflect the arithmetic mean of Likert scale responses on a scale of [-2, +2]; negative values indicate preference for Baseline AI, positive values indicate preference for Wayfinding AI. Error bars represent 95\% confidence intervals determined by bootstrap.}
    \label{fig:sxs_summary}
    \Description{This figure has two panels summarizing responses to questions about which of the two AIs participants preferred. Panel A shows horizontally-stacked bar charts that show the overall distribution of preference along different dimensions.  Counts of responses for each of the 5 options are annotated on each set of bars. The counts for each are (in order): Strongly prefer Wayfinding AI, Somewhat prefer Wayfinding AI, Neutral, Somewhat prefer Baseline AI, Strongly prefer Baseline AI. Raw counts of responses for each axis are marked: Helpfulness: 15, 22, 39, 16, 7; Relevance: 21, 26, 33, 15, 5; Tailoring: 18, 22, 38, 14, 8; Goal understanding: 15, 21, 42, 14, 8; Ease of use: 16, 18, 39, 14, 12; Efficiency: 16, 21, 32, 18, 13; Future search: 21, 28, 18, 19, 14. Panel B shows the mean preference score (treated as a continuous variable) along with 95\% confidence intervals for each dimension. The preference is significantly above 0 (favoring Wayfinding AI) for Helpfulness, Relevance, Tailoring, and Goal understanding.}
\end{figure*}

When asked to explicitly compare the two AIs they tested, participants significantly preferred the Wayfinding AI on the dimensions of helpfulness, relevance of the questions asked, degree of interactions tailored to the participant’s situation, and goal understanding (Figure \ref{fig:sxs_summary}). Responses were neutral for ease of use, efficiency, help in achieving goal and which AI the user would prefer for a future search.

\subsubsection{Different conversation dynamics with the Wayfinding AI}

We observed marked differences in the structure of conversations between the two AIs in Study 4. The initial responses of Baseline AI rarely asked questions to users to proactively gain context on the users' concerns; and when it did it sometimes asked more open-ended questions. (E.g., when prompted for help with workout planning, it completed its response with "Let me know what you're aiming for, and we can build from there!") By comparison, the Wayfinding AI asked specific clarifying questions for 96\% of conversations (Table \ref{tab:clarifying_questions}). For instance, a participant hoping to understand a recent test result was asked "Have you had any recent infections, surgeries or injuries?" (Table \ref{tab:chat-history}).

\begin{table}[ht]
    \centering
    \begin{tabular}{l|c|c}
    \toprule[2pt]
    Questions asked in first agent response & Wayfinding AI & Baseline AI \\
    \midrule[2pt]
    No questions & 4.2\% & 90.1\% \\
    Open-ended question(s) & 8.4\% & 1.8\% \\
    Clarifying question(s) & 87.7\% & 8.0\% \\
    \bottomrule[2pt]
    \end{tabular}
    \caption{Breakdown of first responses by each AI in Study 4 based on the type of questions it asked of users. The visualization for question types for 13 turns are presented in Appendix Figure \ref{fig:chatbot_response_type}.}
    \label{tab:clarifying_questions}
\end{table}

Compared to the Baseline AI, conversations with the Wayfinding AI tended to be longer (Mean 4.51 turns for Wayfinding AI vs. 3.69 turns for Baseline AI; p = 0.025, paired T test). This difference was driven by substantially longer conversations for COSI questions (p < 0.001), while non-COSI questions were not significantly different in length (p = 0.105; Table \ref{tab:conversation_lengths}).

\begin{table}[ht]
    \centering
    \begin{tabular}{m{3.3cm}|m{2cm}|m{2cm}|m{1.5cm}|m{1.8cm}|m{2cm}}
    \toprule[2pt]
 &
 \textbf{Number of conversation turns, Wayfinding AI} &
 \textbf{Number of conversation turns, \makecell[l]{Baseline AI}} &
 \textbf{N, Wayfinding AI} &
 \textbf{N, \makecell[l]{Baseline AI}} &
 \textbf{Difference in conversation length, p (Paired T test)}  \\

 \midrule[2pt]

All data &
4.51
[3.99, 5.03] &
3.69
[3.22, 4.15] &
119 &
112 &
0.0252 \\
\hline

Cause-of-symptom-intent topic (COSI) &
4.96
[4.32, 5.61] &
3.29
[2.79, 3.78] &
84 &
73 &
0.0009 \\
\hline

Non-cause-of-symptom-intent topic (Non-COSI) &
3.43
[2.67, 4.19] &
4.44
[3.52, 5.35] &
35 &
39 &
0.1050 \\

 \bottomrule[0.05cm]

    \end{tabular}
    \caption{Summary of conversation lengths, for all conversations, conversations with a cause-of-symptom intent (COSI), and conversations with a non-cause-of-symptom intent (Non-COSI). Metrics include the point estimate followed by the 95\% confidence interval determined by bootstrap. Note that N reflects the number of participants (out of 130 total) who opted to upload a copy of their conversation after interacting with the Wayfinding AI and Baseline AI respectively. Conversation upload was an optional step in the study flow.}
    \label{tab:conversation_lengths}
\end{table}

\begin{figure*}[h]
    \centering
    \includegraphics[width=\textwidth]{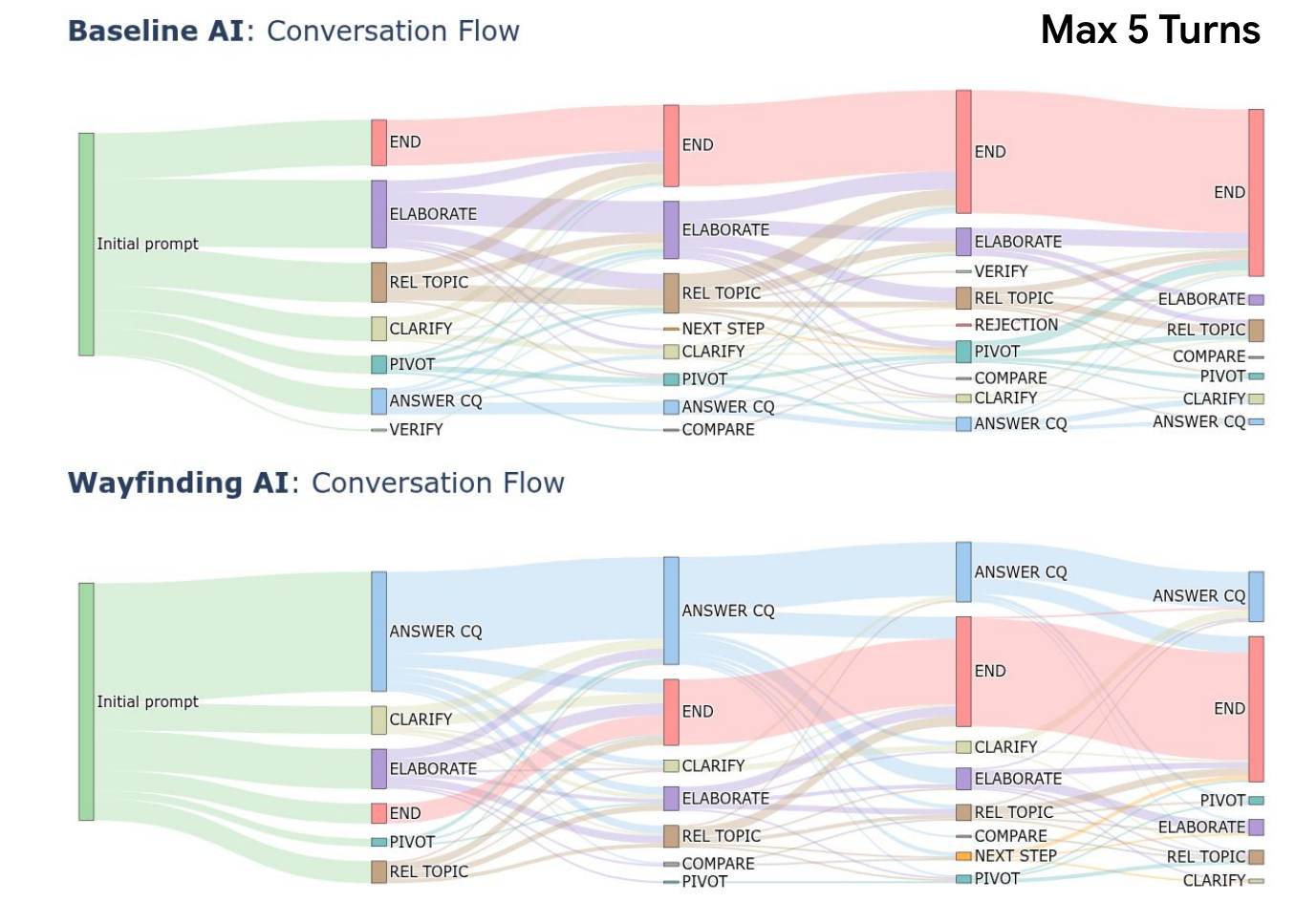}
    \caption{Sankey flow diagrams illustrating distributions of user prompt types across conversations in Study 4. Top, summary of all conversations with Baseline AI. Bottom, summary of all conversations with Wayfinding AI. Nodes indicate different user prompt types. Thickness of lines connecting two nodes indicates relative proportion of conversations with the first prompt type followed by the second prompt type. The first 5 turns are shown. The abbreviations describing each turn type are described in more detail in the Appendix.
    Abbreviations: ANSWER\_CQ: Answering clarifying questions asked by the AI; ELABORATION: Request for AI to elaborate on an aspect of its response; REL\_TOPIC: Related topic exploration; CLARIFY: user provides refinement or clarification of their need without prompting; PIVOT: User pivots to a new topic; COMPARE: user asks AI to compare two entities related to the question; NEXT\_STEP: User asks about a task-related next step; VERIFY: User expresses uncertainty or seeks confirmation of information provided. Details on how labels were applied to conversations are in the Appendix.}
    \label{fig:sankey_diagram}
    \Description{This shows Sankey diagram of user turns across the first 5 turns of conversations for each AI. Each turn is shown as a vertically-arranged set of bars, whose height reflects the proportion of that turn of each type; colored areas connect each bar with the bar in the next turn in proportion to the number of conversations with turns of that type. At the top is the chart for the Baseline AI: the most common prompt types are to End the conversation, or ask the AI to Elaborate. At the bottom is the chart for the Wayfinding AI: The most common prompt types are to Answer clarifying questions, and to or ask the AI to Elaborate. Over turns, more conversations end earlier for the Baseline AI, whereas many conversations in the Wayfinding AI involve sequential answering of clarifying questions.}
\end{figure*}

We classified each user prompt across turns in their conversations (details in Appendix). The patterns of conversations were sharply different between AIs (Figure \ref{fig:sankey_diagram}). With the Wayfinding AI, users engaged in a much higher rate of responding to AI responses with clarifying information (e.g. "I enjoy walking" when asked what type of activities the user enjoyed, for the question of how often to work out; or "I have painful itching" when asked whether a rash has any symptoms.). Conversely, when using Baseline AI, users were much more likely to elaborate, meaning they volunteered information that was not directly in response to the AI response (e.g. "what sleep medications are non habit forming?" when asking about insomnia treatments and being provided a set of recommendations); and were relatively more likely to pivot to different topics (e.g. "Can you reverse type 2 diabetes?" after initially asking "What is the prevnar vaccine? What does it protect you from?") and/or end the conversation earlier.

\newpage
\section{Discussion}

In this work, we provide qualitative and quantitative evidence across four studies demonstrating that people value AIs that explicitly engage in context-seeking to better understand their health questions. Such ``Wayfinding'' AIs result in longer conversations with more turns, that are preferred over conversations with standard AIs, and are perceived as being more relevant to the person’s individual circumstances.

These findings suggest that design considerations in AI behavior and interface can dramatically impact the quality of health information obtained. This also has implications for how AIs may be developed and assessed, and how this technology may impact access to and quality of health information.

\subsection{AI design can affect conversational dynamics}

We show that design choices in AIs can result in significant user-perceived differences in the quality of conversations. Developing an AI that proactively engages in context-seeking results in more relevant information being provided, which is overall viewed by participants as more helpful and tailored to their goals and circumstances (Figure \ref{fig:sxs_summary}). Conversations are also longer and more detailed (Table \ref{tab:conversation_lengths} and Figure \ref{fig:sankey_diagram}).

Design choices include several aspects of AI interface and behavior, which may interact. AI response length can often be long, and length is known to influence human judgments of answer quality \cite{sun2019compare, liu2022revisiting, xu2023critical, li2024does}. We saw in Study 2 that this length also interacts with context-seeking: Participants mostly ignored the clarifying questions the Wayfinding AI from that study asked, in large part because of long responses that did not consistently place clarifying questions in the same location. AIs that have been further trained to ask questions in a consistent format, and UI affordances to highlight clarifying questions, can restore user engagement with context-seeking (as seen in Studies 3 and 4). 

Proactive context-seeking results in behavior that is more in line with human experts, who when conversing with laypeople on topics will often "zoom in" to better understand details \cite{freeman2025zoom} often with targeted clarifying questions \cite{li2024conversational, li2025comparative, bromme2005barriers}. Conversational patterns between laypeople and either an LLM or a clinician presented as an LLM have been shown to systematically differ in terms of how information is provided, and non-need related communication such as empathy \cite{li2024conversational}. Further development of LLMs may aid in other conversational dynamics that human experts tend to engage in, such as "zooming out" to understand wider goals and "reframing" when a layperson is operating on an unsupported premise \cite{freeman2025zoom}.

\subsection{Measuring the value of context-seeking}

Our work focuses on laypeople using AI to understand their own real-world health questions and sharing their impressions of their experience. This approach is distinct from most current approaches for assessing AI performance on health topics. Existing benchmarks focus primarily on closed-form question sets, such as those given to medical students \cite{singhal2023large, nori2023capabilities}. Evaluations of model performance also assess how well they answer open-ended questions, including questions commonly asked online \cite{singhal2025toward, pfohl2024toolbox, bedi2025medhelm}; but these have tended to focus on single-turn responses \cite{hao2025retrospective}. More recent work has started exploring multi-turn evaluation, either by simulating interactive conversations \cite{li2024mediq} or providing clinical review of responses for one turn in a conversation \cite{arora2025healthbench}. 

These automated approaches to measuring health information may overestimate the ability of models to help users, by assuming they know what relevant context to share \cite{bean2025clinical}. Work assessing context-seeking in LLMs found that directly prompting LLMs to ask clarifying questions unexpectedly degraded performance \cite{li2024mediq}, suggesting both that the development of effective context-seeking is non-trivial, and that existing benchmarks may not fully assess the value of perceived relevance and tailoring to users. More broadly, most evaluation approaches rely on assessments of interactions by human raters, by rubrics designed by human raters, or by automated evaluation methods. The extent to which these approaches result in better alignment with the preferences of the people asking the questions (versus a rater assessing the conversation after-the-fact) is unclear.

\subsection{Limitations \& Future Work}

This work has some limitations. Our study focused on text-only AIs; many real-world health questions may be multi-modal, involving images, audio, or video. The set of participants was also limited to the US and these findings may not reflect sentiments from people outside the US. Participants conversed with AIs twice, as part of a study; this behavior may not precisely match behavior if participants were using either AI when they first had their question. The AIs tested here varied both in context-seeking behavior and in UI presentation (motivated by the findings of Studies 2 and 3 which suggested the UI may aid in context seeking by separating out clarifying questions). Our findings therefore reflect the aggregate effects of both UI and model behavior. Future research will be needed to better understand the separate contributions of UI and model behavior, and investigate context-seeking in a multi-modal context, and with a broader sample of participants.

\section{Conclusions}

AI technology, in distilling a large knowledge corpus and tailoring it to individual users, may greatly broaden layperson access to expertise. Within the health domain, this has the promise to expand access to high-quality medical information. Yet it also poses many risks, including that people will be misinformed, overwhelmed, or confused. The value people extract from this technology, and their ability to apply it to their own circumstances, will depend acutely on a range of design considerations. There is an urgency in elucidating these principles, and applying it to develop accurate and accessible tools. 
Our work contributes to this effort by demonstrating the impact of context-seeking on conversational health AIs. By encouraging proactive collection of information, the pattern and quality of interactions with an AI can vary significantly. The resulting conversations are viewed as more helpful, relevant, and tailored to each individual's circumstances. Future work is needed to understand how these different patterns of information-seeking may inform decisions around seeking care, and ultimately whether they can promote better health outcomes.

\begin{acks}
We thank Laura Vardoulakis, Tiffany Guo, and Meredith Ringel Morris for their thoughtful reviews of the manuscript.
\end{acks}

\bibliographystyle{ACM-Reference-Format}
\bibliography{references}


\begin{thebibliography}{74}


\ifx \showCODEN    \undefined \def \showCODEN     #1{\unskip}     \fi
\ifx \showISBNx    \undefined \def \showISBNx     #1{\unskip}     \fi
\ifx \showISBNxiii \undefined \def \showISBNxiii  #1{\unskip}     \fi
\ifx \showISSN     \undefined \def \showISSN      #1{\unskip}     \fi
\ifx \showLCCN     \undefined \def \showLCCN      #1{\unskip}     \fi
\ifx \shownote     \undefined \def \shownote      #1{#1}          \fi
\ifx \showarticletitle \undefined \def \showarticletitle #1{#1}   \fi
\ifx \showURL      \undefined \def \showURL       {\relax}        \fi
\providecommand\bibfield[2]{#2}
\providecommand\bibinfo[2]{#2}
\providecommand\natexlab[1]{#1}
\providecommand\showeprint[2][]{arXiv:#2}

\bibitem[Arora et~al\mbox{.}(2025)]%
        {arora2025healthbench}
\bibfield{author}{\bibinfo{person}{Rahul~K Arora}, \bibinfo{person}{Jason Wei}, \bibinfo{person}{Rebecca~Soskin Hicks}, \bibinfo{person}{Preston Bowman}, \bibinfo{person}{Joaquin Qui{\~n}onero-Candela}, \bibinfo{person}{Foivos Tsimpourlas}, \bibinfo{person}{Michael Sharman}, \bibinfo{person}{Meghan Shah}, \bibinfo{person}{Andrea Vallone}, \bibinfo{person}{Alex Beutel}, {et~al\mbox{.}}} \bibinfo{year}{2025}\natexlab{}.
\newblock \showarticletitle{Healthbench: Evaluating large language models towards improved human health}.
\newblock \bibinfo{journal}{\emph{arXiv preprint arXiv:2505.08775}} (\bibinfo{year}{2025}).
\newblock


\bibitem[Augustaitis et~al\mbox{.}(2021)]%
        {augustaitis2021transgender}
\bibfield{author}{\bibinfo{person}{Laima Augustaitis}, \bibinfo{person}{Leland~A. Merrill}, \bibinfo{person}{Kristi~E Gamarel}, {and} \bibinfo{person}{Oliver~L. Haimson}.} \bibinfo{year}{2021}\natexlab{}.
\newblock \showarticletitle{Online Transgender Health Information Seeking: Facilitators, Barriers, and Future Directions}. In \bibinfo{booktitle}{\emph{Proceedings of the 2021 CHI Conference on Human Factors in Computing Systems}} (Yokohama, Japan) \emph{(\bibinfo{series}{CHI '21})}. \bibinfo{publisher}{Association for Computing Machinery}, \bibinfo{address}{New York, NY, USA}, Article \bibinfo{articleno}{205}, \bibinfo{numpages}{14}~pages.
\newblock
\showISBNx{9781450380966}
\href{https://doi.org/10.1145/3411764.3445091}{doi:\nolinkurl{10.1145/3411764.3445091}}


\bibitem[Ayre et~al\mbox{.}(2025)]%
        {ayre2025use}
\bibfield{author}{\bibinfo{person}{Julie Ayre}, \bibinfo{person}{Erin Cvejic}, {and} \bibinfo{person}{Kirsten~J McCaffery}.} \bibinfo{year}{2025}\natexlab{}.
\newblock \showarticletitle{Use of ChatGPT to obtain health information in Australia, 2024: insights from a nationally representative survey}.
\newblock \bibinfo{journal}{\emph{Medical Journal of Australia}} \bibinfo{volume}{222}, \bibinfo{number}{4} (\bibinfo{year}{2025}), \bibinfo{pages}{210--212}.
\newblock


\bibitem[Babic et~al\mbox{.}(2021)]%
        {babic2021direct}
\bibfield{author}{\bibinfo{person}{Boris Babic}, \bibinfo{person}{Sara Gerke}, \bibinfo{person}{Theodoros Evgeniou}, {and} \bibinfo{person}{I~Glenn Cohen}.} \bibinfo{year}{2021}\natexlab{}.
\newblock \showarticletitle{Direct-to-consumer medical machine learning and artificial intelligence applications}.
\newblock \bibinfo{journal}{\emph{Nature Machine Intelligence}} \bibinfo{volume}{3}, \bibinfo{number}{4} (\bibinfo{year}{2021}), \bibinfo{pages}{283--287}.
\newblock


\bibitem[Bean et~al\mbox{.}(2025)]%
        {bean2025clinical}
\bibfield{author}{\bibinfo{person}{Andrew~M Bean}, \bibinfo{person}{Rebecca Payne}, \bibinfo{person}{Guy Parsons}, \bibinfo{person}{Hannah~Rose Kirk}, \bibinfo{person}{Juan Ciro}, \bibinfo{person}{Rafael Mosquera}, \bibinfo{person}{Sara~Hincapi{\'e} Monsalve}, \bibinfo{person}{Aruna~S Ekanayaka}, \bibinfo{person}{Lionel Tarassenko}, \bibinfo{person}{Luc Rocher}, {et~al\mbox{.}}} \bibinfo{year}{2025}\natexlab{}.
\newblock \showarticletitle{Clinical knowledge in LLMs does not translate to human interactions}.
\newblock \bibinfo{journal}{\emph{arXiv preprint arXiv:2504.18919}} (\bibinfo{year}{2025}).
\newblock


\bibitem[Becker(2004)]%
        {becker2004study}
\bibfield{author}{\bibinfo{person}{Shirley~Ann Becker}.} \bibinfo{year}{2004}\natexlab{}.
\newblock \showarticletitle{A study of web usability for older adults seeking online health resources}.
\newblock \bibinfo{journal}{\emph{ACM Trans. Comput.-Hum. Interact.}} \bibinfo{volume}{11}, \bibinfo{number}{4} (\bibinfo{date}{Dec.} \bibinfo{year}{2004}), \bibinfo{pages}{387–406}.
\newblock
\showISSN{1073-0516}
\href{https://doi.org/10.1145/1035575.1035578}{doi:\nolinkurl{10.1145/1035575.1035578}}


\bibitem[Bedi et~al\mbox{.}(2025)]%
        {bedi2025medhelm}
\bibfield{author}{\bibinfo{person}{Suhana Bedi}, \bibinfo{person}{Hejie Cui}, \bibinfo{person}{Miguel Fuentes}, \bibinfo{person}{Alyssa Unell}, \bibinfo{person}{Michael Wornow}, \bibinfo{person}{Juan~M Banda}, \bibinfo{person}{Nikesh Kotecha}, \bibinfo{person}{Timothy Keyes}, \bibinfo{person}{Yifan Mai}, \bibinfo{person}{Mert Oez}, {et~al\mbox{.}}} \bibinfo{year}{2025}\natexlab{}.
\newblock \showarticletitle{MedHELM: Holistic Evaluation of Large Language Models for Medical Tasks}.
\newblock \bibinfo{journal}{\emph{arXiv preprint arXiv:2505.23802}} (\bibinfo{year}{2025}).
\newblock


\bibitem[Bromme et~al\mbox{.}(2005)]%
        {bromme2005barriers}
\bibfield{author}{\bibinfo{person}{Rainer Bromme}, \bibinfo{person}{Regina Jucks}, {and} \bibinfo{person}{Anne Runde}.} \bibinfo{year}{2005}\natexlab{}.
\newblock \showarticletitle{Barriers and biases in computer-mediated expert-layperson-communication: An overview and insights into the field of medical advice}.
\newblock \bibinfo{journal}{\emph{Barriers and biases in computer-mediated knowledge communication: And how they may be overcome}} (\bibinfo{year}{2005}), \bibinfo{pages}{89--118}.
\newblock


\bibitem[Chan et~al\mbox{.}(2025)]%
        {chan2025odyssey}
\bibfield{author}{\bibinfo{person}{Ka~I Chan}, \bibinfo{person}{Siying Hu}, \bibinfo{person}{Yuntao Wang}, \bibinfo{person}{Xuhai Xu}, \bibinfo{person}{Zhicong Lu}, {and} \bibinfo{person}{Yuanchun Shi}.} \bibinfo{year}{2025}\natexlab{}.
\newblock \showarticletitle{The Odyssey Journey: Top-Tier Medical Resource Seeking for Specialized Disorder in China}. In \bibinfo{booktitle}{\emph{Proceedings of the 2025 CHI Conference on Human Factors in Computing Systems}} \emph{(\bibinfo{series}{CHI '25})}. \bibinfo{publisher}{Association for Computing Machinery}, \bibinfo{address}{New York, NY, USA}, Article \bibinfo{articleno}{1048}, \bibinfo{numpages}{18}~pages.
\newblock
\showISBNx{9798400713941}
\href{https://doi.org/10.1145/3706598.3713315}{doi:\nolinkurl{10.1145/3706598.3713315}}


\bibitem[Chang et~al\mbox{.}(2022)]%
        {chang2022would}
\bibfield{author}{\bibinfo{person}{I-Chiu Chang}, \bibinfo{person}{Yi-Syuan Shih}, {and} \bibinfo{person}{Kuang-Ming Kuo}.} \bibinfo{year}{2022}\natexlab{}.
\newblock \showarticletitle{Why would you use medical chatbots? interview and survey}.
\newblock \bibinfo{journal}{\emph{International Journal of Medical Informatics}}  \bibinfo{volume}{165} (\bibinfo{year}{2022}), \bibinfo{pages}{104827}.
\newblock


\bibitem[Clarke and Braun(2014)]%
        {clarke2014thematic}
\bibfield{author}{\bibinfo{person}{Victoria Clarke} {and} \bibinfo{person}{Virginia Braun}.} \bibinfo{year}{2014}\natexlab{}.
\newblock \showarticletitle{Thematic analysis}.
\newblock In \bibinfo{booktitle}{\emph{Encyclopedia of critical psychology}}. \bibinfo{publisher}{Springer}, \bibinfo{pages}{1947--1952}.
\newblock


\bibitem[Comanici et~al\mbox{.}(2025)]%
        {comanici2025gemini}
\bibfield{author}{\bibinfo{person}{Gheorghe Comanici}, \bibinfo{person}{Eric Bieber}, \bibinfo{person}{Mike Schaekermann}, \bibinfo{person}{Ice Pasupat}, \bibinfo{person}{Noveen Sachdeva}, \bibinfo{person}{Inderjit Dhillon}, \bibinfo{person}{Marcel Blistein}, \bibinfo{person}{Ori Ram}, \bibinfo{person}{Dan Zhang}, \bibinfo{person}{Evan Rosen}, {et~al\mbox{.}}} \bibinfo{year}{2025}\natexlab{}.
\newblock \showarticletitle{Gemini 2.5: Pushing the Frontier with Advanced Reasoning, Multimodality, Long Context, and Next Generation Agentic Capabilities}.
\newblock \bibinfo{journal}{\emph{arXiv preprint arXiv:2507.06261}} (\bibinfo{year}{2025}).
\newblock


\bibitem[De~Choudhury et~al\mbox{.}(2014)]%
        {choudhyry2014seeking}
\bibfield{author}{\bibinfo{person}{Munmun De~Choudhury}, \bibinfo{person}{Meredith~Ringel Morris}, {and} \bibinfo{person}{Ryen~W. White}.} \bibinfo{year}{2014}\natexlab{}.
\newblock \showarticletitle{Seeking and sharing health information online: comparing search engines and social media}. In \bibinfo{booktitle}{\emph{Proceedings of the SIGCHI Conference on Human Factors in Computing Systems}} (Toronto, Ontario, Canada) \emph{(\bibinfo{series}{CHI '14})}. \bibinfo{publisher}{Association for Computing Machinery}, \bibinfo{address}{New York, NY, USA}, \bibinfo{pages}{1365–1376}.
\newblock
\showISBNx{9781450324731}
\href{https://doi.org/10.1145/2556288.2557214}{doi:\nolinkurl{10.1145/2556288.2557214}}


\bibitem[De~Freitas and Cohen(2024)]%
        {de2024health}
\bibfield{author}{\bibinfo{person}{Julian De~Freitas} {and} \bibinfo{person}{I~Glenn Cohen}.} \bibinfo{year}{2024}\natexlab{}.
\newblock \showarticletitle{The health risks of generative AI-based wellness apps}.
\newblock \bibinfo{journal}{\emph{Nature medicine}} \bibinfo{volume}{30}, \bibinfo{number}{5} (\bibinfo{year}{2024}), \bibinfo{pages}{1269--1275}.
\newblock


\bibitem[Detjen et~al\mbox{.}(2025)]%
        {detjen2025trusted}
\bibfield{author}{\bibinfo{person}{Henrik~HJ Detjen}, \bibinfo{person}{Lars Densky}, \bibinfo{person}{Niklas von Kalckreuth}, {and} \bibinfo{person}{Marvin Kopka}.} \bibinfo{year}{2025}\natexlab{}.
\newblock \showarticletitle{Who is Trusted for a Second Opinion? Comparing Collective Advice from a Medical AI and Physicians in Biopsy Decisions After Mammography Screening}. In \bibinfo{booktitle}{\emph{Proceedings of the 2025 CHI Conference on Human Factors in Computing Systems}}. \bibinfo{pages}{1--15}.
\newblock


\bibitem[Dewan et~al\mbox{.}(2024)]%
        {dewan2024teen}
\bibfield{author}{\bibinfo{person}{Umama Dewan}, \bibinfo{person}{Cora Sula}, {and} \bibinfo{person}{Nora Mcdonald}.} \bibinfo{year}{2024}\natexlab{}.
\newblock \showarticletitle{Teen Reproductive Health Information Seeking and Sharing Post-Roe}. In \bibinfo{booktitle}{\emph{Proceedings of the 2024 CHI Conference on Human Factors in Computing Systems}} (Honolulu, HI, USA) \emph{(\bibinfo{series}{CHI '24})}. \bibinfo{publisher}{Association for Computing Machinery}, \bibinfo{address}{New York, NY, USA}, Article \bibinfo{articleno}{814}, \bibinfo{numpages}{12}~pages.
\newblock
\showISBNx{9798400703300}
\href{https://doi.org/10.1145/3613904.3641934}{doi:\nolinkurl{10.1145/3613904.3641934}}


\bibitem[Di~Novi et~al\mbox{.}(2024)]%
        {di2024online}
\bibfield{author}{\bibinfo{person}{Cinzia Di~Novi}, \bibinfo{person}{Matija Kovacic}, {and} \bibinfo{person}{Cristina~Elisa Orso}.} \bibinfo{year}{2024}\natexlab{}.
\newblock \showarticletitle{Online health information seeking behavior, healthcare access, and health status during exceptional times}.
\newblock \bibinfo{journal}{\emph{Journal of Economic Behavior \& Organization}}  \bibinfo{volume}{220} (\bibinfo{year}{2024}), \bibinfo{pages}{675--690}.
\newblock


\bibitem[Eichenberger et~al\mbox{.}(2025)]%
        {eichenberger2025case}
\bibfield{author}{\bibinfo{person}{Audrey Eichenberger}, \bibinfo{person}{Stephen Thielke}, {and} \bibinfo{person}{Adam Van~Buskirk}.} \bibinfo{year}{2025}\natexlab{}.
\newblock \showarticletitle{A Case of Bromism Influenced by Use of Artificial Intelligence}.
\newblock \bibinfo{journal}{\emph{Annals of Internal Medicine: Clinical Cases}} \bibinfo{volume}{4}, \bibinfo{number}{8} (\bibinfo{year}{2025}), \bibinfo{pages}{e241260}.
\newblock


\bibitem[Engel and Morgan(1973)]%
        {engel1973interviewing}
\bibfield{author}{\bibinfo{person}{George~Libman Engel} {and} \bibinfo{person}{William~L Morgan}.} \bibinfo{year}{1973}\natexlab{}.
\newblock \showarticletitle{Interviewing the patient}.
\newblock \bibinfo{journal}{\emph{(No Title)}} (\bibinfo{year}{1973}).
\newblock


\bibitem[Esteva et~al\mbox{.}(2019)]%
        {esteva2019guide}
\bibfield{author}{\bibinfo{person}{Andre Esteva}, \bibinfo{person}{Alexandre Robicquet}, \bibinfo{person}{Bharath Ramsundar}, \bibinfo{person}{Volodymyr Kuleshov}, \bibinfo{person}{Mark DePristo}, \bibinfo{person}{Katherine Chou}, \bibinfo{person}{Claire Cui}, \bibinfo{person}{Greg Corrado}, \bibinfo{person}{Sebastian Thrun}, {and} \bibinfo{person}{Jeff Dean}.} \bibinfo{year}{2019}\natexlab{}.
\newblock \showarticletitle{A guide to deep learning in healthcare}.
\newblock \bibinfo{journal}{\emph{Nature medicine}} \bibinfo{volume}{25}, \bibinfo{number}{1} (\bibinfo{year}{2019}), \bibinfo{pages}{24--29}.
\newblock


\bibitem[Fan et~al\mbox{.}(2021)]%
        {fan2021utilization}
\bibfield{author}{\bibinfo{person}{Xiangmin Fan}, \bibinfo{person}{Daren Chao}, \bibinfo{person}{Zhan Zhang}, \bibinfo{person}{Dakuo Wang}, \bibinfo{person}{Xiaohua Li}, {and} \bibinfo{person}{Feng Tian}.} \bibinfo{year}{2021}\natexlab{}.
\newblock \showarticletitle{Utilization of self-diagnosis health chatbots in real-world settings: case study}.
\newblock \bibinfo{journal}{\emph{Journal of medical Internet research}} \bibinfo{volume}{23}, \bibinfo{number}{1} (\bibinfo{year}{2021}), \bibinfo{pages}{e19928}.
\newblock


\bibitem[Flugelman(2021)]%
        {flugelman2021history}
\bibfield{author}{\bibinfo{person}{Moshe~Y Flugelman}.} \bibinfo{year}{2021}\natexlab{}.
\newblock \showarticletitle{History-taking revisited: Simple techniques to foster patient collaboration, improve data attainment, and establish trust with the patient}.
\newblock \bibinfo{journal}{\emph{GMS Journal for Medical Education}} \bibinfo{volume}{38}, \bibinfo{number}{6} (\bibinfo{year}{2021}), \bibinfo{pages}{Doc109}.
\newblock


\bibitem[Freeman et~al\mbox{.}(2025)]%
        {freeman2025zoom}
\bibfield{author}{\bibinfo{person}{Beverly Freeman}, \bibinfo{person}{Roma Ruparel}, {and} \bibinfo{person}{Laura~M Vardoulakis}.} \bibinfo{year}{2025}\natexlab{}.
\newblock \showarticletitle{Zoom in, Zoom out, Reframe: Domain Experts’ Strategies for Addressing Non-Experts' Complex Questions}. In \bibinfo{booktitle}{\emph{Proceedings of the Extended Abstracts of the CHI Conference on Human Factors in Computing Systems}}. \bibinfo{pages}{1--7}.
\newblock


\bibitem[Gaube et~al\mbox{.}(2021)]%
        {gaube2021ai}
\bibfield{author}{\bibinfo{person}{Susanne Gaube}, \bibinfo{person}{Harini Suresh}, \bibinfo{person}{Martina Raue}, \bibinfo{person}{Alexander Merritt}, \bibinfo{person}{Seth~J Berkowitz}, \bibinfo{person}{Eva Lermer}, \bibinfo{person}{Joseph~F Coughlin}, \bibinfo{person}{John~V Guttag}, \bibinfo{person}{Errol Colak}, {and} \bibinfo{person}{Marzyeh Ghassemi}.} \bibinfo{year}{2021}\natexlab{}.
\newblock \showarticletitle{Do as AI say: susceptibility in deployment of clinical decision-aids}.
\newblock \bibinfo{journal}{\emph{NPJ digital medicine}} \bibinfo{volume}{4}, \bibinfo{number}{1} (\bibinfo{year}{2021}), \bibinfo{pages}{31}.
\newblock


\bibitem[Goodman et~al\mbox{.}(2023)]%
        {goodman2023accuracy}
\bibfield{author}{\bibinfo{person}{Rachel~S Goodman}, \bibinfo{person}{J~Randall Patrinely}, \bibinfo{person}{Cosby~A Stone}, \bibinfo{person}{Eli Zimmerman}, \bibinfo{person}{Rebecca~R Donald}, \bibinfo{person}{Sam~S Chang}, \bibinfo{person}{Sean~T Berkowitz}, \bibinfo{person}{Avni~P Finn}, \bibinfo{person}{Eiman Jahangir}, \bibinfo{person}{Elizabeth~A Scoville}, {et~al\mbox{.}}} \bibinfo{year}{2023}\natexlab{}.
\newblock \showarticletitle{Accuracy and reliability of chatbot responses to physician questions}.
\newblock \bibinfo{journal}{\emph{JAMA network open}} \bibinfo{volume}{6}, \bibinfo{number}{10} (\bibinfo{year}{2023}), \bibinfo{pages}{e2336483--e2336483}.
\newblock


\bibitem[Hao et~al\mbox{.}(2025a)]%
        {hao2025retrospective}
\bibfield{author}{\bibinfo{person}{Yuexing Hao}, \bibinfo{person}{Jason Holmes}, \bibinfo{person}{Jared Hobson}, \bibinfo{person}{Alexandra Bennett}, \bibinfo{person}{Elizabeth~L McKone}, \bibinfo{person}{Daniel~K Ebner}, \bibinfo{person}{David~M Routman}, \bibinfo{person}{Satomi Shiraishi}, \bibinfo{person}{Samir~H Patel}, \bibinfo{person}{Nathan~Y Yu}, {et~al\mbox{.}}} \bibinfo{year}{2025}\natexlab{a}.
\newblock \showarticletitle{Retrospective Comparative Analysis of Prostate Cancer In-Basket Messages: Responses From Closed-Domain Large Language Models Versus Clinical Teams}.
\newblock \bibinfo{journal}{\emph{Mayo Clinic Proceedings: Digital Health}} \bibinfo{volume}{3}, \bibinfo{number}{1} (\bibinfo{year}{2025}), \bibinfo{pages}{100198}.
\newblock


\bibitem[Hao et~al\mbox{.}(2024)]%
        {hao_advancing_2024}
\bibfield{author}{\bibinfo{person}{Yuexing Hao}, \bibinfo{person}{Zeyu Liu}, \bibinfo{person}{Robert~N Riter}, {and} \bibinfo{person}{Saleh Kalantari}.} \bibinfo{year}{2024}\natexlab{}.
\newblock \showarticletitle{Advancing {Patient}-{Centered} {Shared} {Decision}-{Making} with {AI} {Systems} for {Older} {Adult} {Cancer} {Patients}}. In \bibinfo{booktitle}{\emph{Proceedings of the {CHI} {Conference} on {Human} {Factors} in {Computing} {Systems}}}. \bibinfo{pages}{1--20}.
\newblock


\bibitem[Hao et~al\mbox{.}(2025b)]%
        {hao2025large}
\bibfield{author}{\bibinfo{person}{Yuexing Hao}, \bibinfo{person}{Zhiwen Qiu}, \bibinfo{person}{Jason Holmes}, \bibinfo{person}{Corinna~E L{\"o}ckenhoff}, \bibinfo{person}{Wei Liu}, \bibinfo{person}{Marzyeh Ghassemi}, {and} \bibinfo{person}{Saleh Kalantari}.} \bibinfo{year}{2025}\natexlab{b}.
\newblock \showarticletitle{Large language model integrations in cancer decision-making: a systematic review and meta-analysis}.
\newblock \bibinfo{journal}{\emph{npj Digital Medicine}} \bibinfo{volume}{8}, \bibinfo{number}{1} (\bibinfo{year}{2025}), \bibinfo{pages}{450}.
\newblock


\bibitem[Harrington and Egede(2023)]%
        {harrington2023trust}
\bibfield{author}{\bibinfo{person}{Christina~N. Harrington} {and} \bibinfo{person}{Lisa Egede}.} \bibinfo{year}{2023}\natexlab{}.
\newblock \showarticletitle{Trust, Comfort and Relatability: Understanding Black Older Adults’ Perceptions of Chatbot Design for Health Information Seeking}. In \bibinfo{booktitle}{\emph{Proceedings of the 2023 CHI Conference on Human Factors in Computing Systems}} (Hamburg, Germany) \emph{(\bibinfo{series}{CHI '23})}. \bibinfo{publisher}{Association for Computing Machinery}, \bibinfo{address}{New York, NY, USA}, Article \bibinfo{articleno}{120}, \bibinfo{numpages}{18}~pages.
\newblock
\showISBNx{9781450394215}
\href{https://doi.org/10.1145/3544548.3580719}{doi:\nolinkurl{10.1145/3544548.3580719}}


\bibitem[Harrington et~al\mbox{.}(2022)]%
        {harrington2022blackadults}
\bibfield{author}{\bibinfo{person}{Christina~N. Harrington}, \bibinfo{person}{Radhika Garg}, \bibinfo{person}{Amanda Woodward}, {and} \bibinfo{person}{Dimitri Williams}.} \bibinfo{year}{2022}\natexlab{}.
\newblock \showarticletitle{“It’s Kind of Like Code-Switching”: Black Older Adults’ Experiences with a Voice Assistant for Health Information Seeking}. In \bibinfo{booktitle}{\emph{Proceedings of the 2022 CHI Conference on Human Factors in Computing Systems}} (New Orleans, LA, USA) \emph{(\bibinfo{series}{CHI '22})}. \bibinfo{publisher}{Association for Computing Machinery}, \bibinfo{address}{New York, NY, USA}, Article \bibinfo{articleno}{604}, \bibinfo{numpages}{15}~pages.
\newblock
\showISBNx{9781450391573}
\href{https://doi.org/10.1145/3491102.3501995}{doi:\nolinkurl{10.1145/3491102.3501995}}


\bibitem[Hofmann and Gilmer(2025)]%
        {HofmannGilmer2023Partners}
\bibfield{author}{\bibinfo{person}{Courtney Hofmann} {and} \bibinfo{person}{Holly Gilmer}.} \bibinfo{year}{2025}\natexlab{}.
\newblock \bibinfo{booktitle}{\emph{Partners in Diagnosis: ChatGPT, a Mother’s Intuition, and a Doctor’s Expertise with Courtney Hofmann}}.
\newblock
\urldef\tempurl%
\url{https://ai-podcast.nejm.org/e/partners-in-diagnosis-chatgpt-a-mother-s-intuition-and-a-doctor-s-expertise-with-courtney-hofman-and-dr-holly-gilmer/}
\showURL{%
\tempurl}
\newblock
\shownote{NEJM AI Podcast}.


\bibitem[Jawale et~al\mbox{.}(2024)]%
        {jawale2024human}
\bibfield{author}{\bibinfo{person}{Toshish Jawale}, \bibinfo{person}{Chaitanya Animesh}, \bibinfo{person}{Sekhar Vallath}, \bibinfo{person}{Kartik Talamadupula}, {and} \bibinfo{person}{Larry Heck}.} \bibinfo{year}{2024}\natexlab{}.
\newblock \showarticletitle{Are human conversations special? a large language model perspective}.
\newblock \bibinfo{journal}{\emph{arXiv preprint arXiv:2403.05045}} (\bibinfo{year}{2024}).
\newblock


\bibitem[Jia et~al\mbox{.}(2021)]%
        {jia2021online}
\bibfield{author}{\bibinfo{person}{Xiaoyun Jia}, \bibinfo{person}{Yan Pang}, {and} \bibinfo{person}{Liangni~Sally Liu}.} \bibinfo{year}{2021}\natexlab{}.
\newblock \showarticletitle{Online health information seeking behavior: a systematic review}. In \bibinfo{booktitle}{\emph{Healthcare}}, Vol.~\bibinfo{volume}{9}. MDPI, \bibinfo{pages}{1740}.
\newblock


\bibitem[Johri et~al\mbox{.}(2024)]%
        {johri2024craft}
\bibfield{author}{\bibinfo{person}{Shreya Johri}, \bibinfo{person}{Jaehwan Jeong}, \bibinfo{person}{Benjamin~A Tran}, \bibinfo{person}{Daniel~I Schlessinger}, \bibinfo{person}{Shannon Wongvibulsin}, \bibinfo{person}{Zhuo~Ran Cai}, \bibinfo{person}{Roxana Daneshjou}, {and} \bibinfo{person}{Pranav Rajpurkar}.} \bibinfo{year}{2024}\natexlab{}.
\newblock \showarticletitle{CRAFT-MD: A conversational evaluation framework for comprehensive assessment of clinical LLMs}. In \bibinfo{booktitle}{\emph{AAAI 2024 Spring Symposium on Clinical Foundation Models}}.
\newblock


\bibitem[Krogue et~al\mbox{.}(2024)]%
        {krogue2024searching}
\bibfield{author}{\bibinfo{person}{Justin~D Krogue}, \bibinfo{person}{Rory Sayres}, \bibinfo{person}{Jay Hartford}, \bibinfo{person}{Amit Talreja}, \bibinfo{person}{Pinal Bavishi}, \bibinfo{person}{Natalie Salaets}, \bibinfo{person}{Kimberley Raiford}, \bibinfo{person}{Jay Nayar}, \bibinfo{person}{Rajan Patel}, \bibinfo{person}{Yossi Matias}, {et~al\mbox{.}}} \bibinfo{year}{2024}\natexlab{}.
\newblock \showarticletitle{Searching for Dermatology Information Online using Images vs Text: a Randomized Study}.
\newblock \bibinfo{journal}{\emph{medRxiv}} (\bibinfo{year}{2024}), \bibinfo{pages}{2024--10}.
\newblock


\bibitem[Laban et~al\mbox{.}(2025)]%
        {laban2025llms}
\bibfield{author}{\bibinfo{person}{Philippe Laban}, \bibinfo{person}{Hiroaki Hayashi}, \bibinfo{person}{Yingbo Zhou}, {and} \bibinfo{person}{Jennifer Neville}.} \bibinfo{year}{2025}\natexlab{}.
\newblock \showarticletitle{Llms get lost in multi-turn conversation}.
\newblock \bibinfo{journal}{\emph{arXiv preprint arXiv:2505.06120}} (\bibinfo{year}{2025}).
\newblock


\bibitem[Lambert and Loiselle(2007)]%
        {lambert2007health}
\bibfield{author}{\bibinfo{person}{Sylvie~D Lambert} {and} \bibinfo{person}{Carmen~G Loiselle}.} \bibinfo{year}{2007}\natexlab{}.
\newblock \showarticletitle{Health information—seeking behavior}.
\newblock \bibinfo{journal}{\emph{Qualitative health research}} \bibinfo{volume}{17}, \bibinfo{number}{8} (\bibinfo{year}{2007}), \bibinfo{pages}{1006--1019}.
\newblock


\bibitem[Levine and Mehrotra(2021)]%
        {levine2021assessment}
\bibfield{author}{\bibinfo{person}{David~M Levine} {and} \bibinfo{person}{Ateev Mehrotra}.} \bibinfo{year}{2021}\natexlab{}.
\newblock \showarticletitle{Assessment of diagnosis and triage in validated case vignettes among nonphysicians before and after internet search}.
\newblock \bibinfo{journal}{\emph{JAMA network open}} \bibinfo{volume}{4}, \bibinfo{number}{3} (\bibinfo{year}{2021}), \bibinfo{pages}{e213287--e213287}.
\newblock


\bibitem[Li et~al\mbox{.}(2024c)]%
        {li2024beyond}
\bibfield{author}{\bibinfo{person}{Brenna Li}, \bibinfo{person}{Ofek Gross}, \bibinfo{person}{Noah Crampton}, \bibinfo{person}{Mamta Kapoor}, \bibinfo{person}{Saba Tauseef}, \bibinfo{person}{Mohit Jain}, \bibinfo{person}{Khai~N Truong}, {and} \bibinfo{person}{Alex Mariakakis}.} \bibinfo{year}{2024}\natexlab{c}.
\newblock \showarticletitle{Beyond the Waiting Room: Patient's Perspectives on the Conversational Nuances of Pre-Consultation Chatbots}. In \bibinfo{booktitle}{\emph{Proceedings of the 2024 CHI Conference on Human Factors in Computing Systems}}. \bibinfo{pages}{1--24}.
\newblock


\bibitem[Li et~al\mbox{.}(2023)]%
        {li2023constraints}
\bibfield{author}{\bibinfo{person}{Brenna Li}, \bibinfo{person}{Tetyana Skoropad}, \bibinfo{person}{Puneet Seth}, \bibinfo{person}{Mohit Jain}, \bibinfo{person}{Khai Truong}, {and} \bibinfo{person}{Alex Mariakakis}.} \bibinfo{year}{2023}\natexlab{}.
\newblock \showarticletitle{Constraints and Workarounds to Support Clinical Consultations in Synchronous Text-based Platforms}. In \bibinfo{booktitle}{\emph{Proceedings of the 2023 CHI Conference on Human Factors in Computing Systems}}. \bibinfo{pages}{1--17}.
\newblock


\bibitem[Li et~al\mbox{.}(2025b)]%
        {li2025comparative}
\bibfield{author}{\bibinfo{person}{Brenna Li}, \bibinfo{person}{Saba Tauseef}, \bibinfo{person}{Khai~N Truong}, {and} \bibinfo{person}{Alex Mariakakis}.} \bibinfo{year}{2025}\natexlab{b}.
\newblock \showarticletitle{A Comparative Analysis of Information Gathering by Chatbots, Questionnaires, and Humans in Clinical Pre-Consultation}. In \bibinfo{booktitle}{\emph{Proceedings of the 2025 CHI Conference on Human Factors in Computing Systems}}. \bibinfo{pages}{1--17}.
\newblock


\bibitem[Li et~al\mbox{.}(2024d)]%
        {li2024conversational}
\bibfield{author}{\bibinfo{person}{Brenna Li}, \bibinfo{person}{Amy Wang}, \bibinfo{person}{Patricia Strachan}, \bibinfo{person}{Julie~Anne S{\'e}guin}, \bibinfo{person}{Sami Lachgar}, \bibinfo{person}{Karyn~C Schroeder}, \bibinfo{person}{Mathias~S Fleck}, \bibinfo{person}{Renee Wong}, \bibinfo{person}{Alan Karthikesalingam}, \bibinfo{person}{Vivek Natarajan}, {et~al\mbox{.}}} \bibinfo{year}{2024}\natexlab{d}.
\newblock \showarticletitle{Conversational AI in health: Design considerations from a Wizard-of-Oz dermatology case study with users, clinicians and a medical LLM}. In \bibinfo{booktitle}{\emph{Extended Abstracts of the CHI Conference on Human Factors in Computing Systems}}. \bibinfo{pages}{1--10}.
\newblock


\bibitem[Li et~al\mbox{.}(2024b)]%
        {li2024mediq}
\bibfield{author}{\bibinfo{person}{Stella Li}, \bibinfo{person}{Vidhisha Balachandran}, \bibinfo{person}{Shangbin Feng}, \bibinfo{person}{Jonathan Ilgen}, \bibinfo{person}{Emma Pierson}, \bibinfo{person}{Pang Wei~W Koh}, {and} \bibinfo{person}{Yulia Tsvetkov}.} \bibinfo{year}{2024}\natexlab{b}.
\newblock \showarticletitle{Mediq: Question-asking llms and a benchmark for reliable interactive clinical reasoning}.
\newblock \bibinfo{journal}{\emph{Advances in Neural Information Processing Systems}}  \bibinfo{volume}{37} (\bibinfo{year}{2024}), \bibinfo{pages}{28858--28888}.
\newblock


\bibitem[Li et~al\mbox{.}(2025a)]%
        {li2025alfaaligningllmsask}
\bibfield{author}{\bibinfo{person}{Shuyue~Stella Li}, \bibinfo{person}{Jimin Mun}, \bibinfo{person}{Faeze Brahman}, \bibinfo{person}{Pedram Hosseini}, \bibinfo{person}{Bryceton~G. Thomas}, \bibinfo{person}{Jessica~M. Sin}, \bibinfo{person}{Bing Ren}, \bibinfo{person}{Jonathan~S. Ilgen}, \bibinfo{person}{Yulia Tsvetkov}, {and} \bibinfo{person}{Maarten Sap}.} \bibinfo{year}{2025}\natexlab{a}.
\newblock \bibinfo{title}{ALFA: Aligning LLMs to Ask Good Questions A Case Study in Clinical Reasoning}.
\newblock
\showeprint[arxiv]{2502.14860}~[cs.CL]
\urldef\tempurl%
\url{https://arxiv.org/abs/2502.14860}
\showURL{%
\tempurl}


\bibitem[Li et~al\mbox{.}(2024a)]%
        {li2024does}
\bibfield{author}{\bibinfo{person}{Tianle Li}, \bibinfo{person}{Anastasios Angelopoulos}, {and} \bibinfo{person}{Wei-Lin Chiang}.} \bibinfo{year}{2024}\natexlab{a}.
\newblock \showarticletitle{Does style matter? disentangling style and substance in chatbot arena}.
\newblock \bibinfo{journal}{\emph{LMSYS Blog}} (\bibinfo{year}{2024}).
\newblock


\bibitem[Liao and Fu(2014)]%
        {liao2014age}
\bibfield{author}{\bibinfo{person}{Q.~Vera Liao} {and} \bibinfo{person}{Wai-Tat Fu}.} \bibinfo{year}{2014}\natexlab{}.
\newblock \showarticletitle{Age differences in credibility judgments of online health information}.
\newblock \bibinfo{journal}{\emph{ACM Trans. Comput.-Hum. Interact.}} \bibinfo{volume}{21}, \bibinfo{number}{1}, Article \bibinfo{articleno}{2} (\bibinfo{date}{Feb.} \bibinfo{year}{2014}), \bibinfo{numpages}{23}~pages.
\newblock
\showISSN{1073-0516}
\href{https://doi.org/10.1145/2534410}{doi:\nolinkurl{10.1145/2534410}}


\bibitem[Liu et~al\mbox{.}(2022)]%
        {liu2022revisiting}
\bibfield{author}{\bibinfo{person}{Yixin Liu}, \bibinfo{person}{Alexander~R Fabbri}, \bibinfo{person}{Pengfei Liu}, \bibinfo{person}{Yilun Zhao}, \bibinfo{person}{Linyong Nan}, \bibinfo{person}{Ruilin Han}, \bibinfo{person}{Simeng Han}, \bibinfo{person}{Shafiq Joty}, \bibinfo{person}{Chien-Sheng Wu}, \bibinfo{person}{Caiming Xiong}, {et~al\mbox{.}}} \bibinfo{year}{2022}\natexlab{}.
\newblock \showarticletitle{Revisiting the gold standard: Grounding summarization evaluation with robust human evaluation}.
\newblock \bibinfo{journal}{\emph{arXiv preprint arXiv:2212.07981}} (\bibinfo{year}{2022}).
\newblock


\bibitem[McDuff et~al\mbox{.}(2025)]%
        {mcduff2025towards}
\bibfield{author}{\bibinfo{person}{Daniel McDuff}, \bibinfo{person}{Mike Schaekermann}, \bibinfo{person}{Tao Tu}, \bibinfo{person}{Anil Palepu}, \bibinfo{person}{Amy Wang}, \bibinfo{person}{Jake Garrison}, \bibinfo{person}{Karan Singhal}, \bibinfo{person}{Yash Sharma}, \bibinfo{person}{Shekoofeh Azizi}, \bibinfo{person}{Kavita Kulkarni}, {et~al\mbox{.}}} \bibinfo{year}{2025}\natexlab{}.
\newblock \showarticletitle{Towards accurate differential diagnosis with large language models}.
\newblock \bibinfo{journal}{\emph{Nature}} (\bibinfo{year}{2025}), \bibinfo{pages}{1--7}.
\newblock


\bibitem[McKinney et~al\mbox{.}(2020)]%
        {mckinney2020international}
\bibfield{author}{\bibinfo{person}{Scott~Mayer McKinney}, \bibinfo{person}{Marcin Sieniek}, \bibinfo{person}{Varun Godbole}, \bibinfo{person}{Jonathan Godwin}, \bibinfo{person}{Natasha Antropova}, \bibinfo{person}{Hutan Ashrafian}, \bibinfo{person}{Trevor Back}, \bibinfo{person}{Mary Chesus}, \bibinfo{person}{Greg~S Corrado}, \bibinfo{person}{Ara Darzi}, {et~al\mbox{.}}} \bibinfo{year}{2020}\natexlab{}.
\newblock \showarticletitle{International evaluation of an AI system for breast cancer screening}.
\newblock \bibinfo{journal}{\emph{Nature}} \bibinfo{volume}{577}, \bibinfo{number}{7788} (\bibinfo{year}{2020}), \bibinfo{pages}{89--94}.
\newblock


\bibitem[Mendel et~al\mbox{.}(2025)]%
        {mendel2025laypeople}
\bibfield{author}{\bibinfo{person}{Tamir Mendel}, \bibinfo{person}{Nina Singh}, \bibinfo{person}{Devin~M Mann}, \bibinfo{person}{Batia Wiesenfeld}, {and} \bibinfo{person}{Oded Nov}.} \bibinfo{year}{2025}\natexlab{}.
\newblock \showarticletitle{Laypeople's Use of and Attitudes Toward Large Language Models and Search Engines for Health Queries: Survey Study}.
\newblock \bibinfo{journal}{\emph{J Med Internet Res}}  \bibinfo{volume}{27} (\bibinfo{date}{13 Feb} \bibinfo{year}{2025}), \bibinfo{pages}{e64290}.
\newblock
\showISSN{1438-8871}
\href{https://doi.org/10.2196/64290}{doi:\nolinkurl{10.2196/64290}}


\bibitem[Milton et~al\mbox{.}(2024)]%
        {milton2024seeking}
\bibfield{author}{\bibinfo{person}{Ashlee Milton}, \bibinfo{person}{Juan~F. Maestre}, \bibinfo{person}{Abhishek Roy}, \bibinfo{person}{Rebecca Umbach}, {and} \bibinfo{person}{Stevie Chancellor}.} \bibinfo{year}{2024}\natexlab{}.
\newblock \showarticletitle{Seeking in Cycles: How Users Leverage Personal Information Ecosystems to Find Mental Health Information}. In \bibinfo{booktitle}{\emph{Proceedings of the 2024 CHI Conference on Human Factors in Computing Systems}} (Honolulu, HI, USA) \emph{(\bibinfo{series}{CHI '24})}. \bibinfo{publisher}{Association for Computing Machinery}, \bibinfo{address}{New York, NY, USA}, Article \bibinfo{articleno}{540}, \bibinfo{numpages}{16}~pages.
\newblock
\showISBNx{9798400703300}
\href{https://doi.org/10.1145/3613904.3641894}{doi:\nolinkurl{10.1145/3613904.3641894}}


\bibitem[Nadarzynski et~al\mbox{.}(2019)]%
        {nadarzynski2019acceptability}
\bibfield{author}{\bibinfo{person}{Tom Nadarzynski}, \bibinfo{person}{Oliver Miles}, \bibinfo{person}{Aimee Cowie}, {and} \bibinfo{person}{Damien Ridge}.} \bibinfo{year}{2019}\natexlab{}.
\newblock \showarticletitle{Acceptability of artificial intelligence (AI)-led chatbot services in healthcare: A mixed-methods study}.
\newblock \bibinfo{journal}{\emph{Digital health}}  \bibinfo{volume}{5} (\bibinfo{year}{2019}), \bibinfo{pages}{2055207619871808}.
\newblock


\bibitem[Nori et~al\mbox{.}(2023)]%
        {nori2023capabilities}
\bibfield{author}{\bibinfo{person}{Harsha Nori}, \bibinfo{person}{Nicholas King}, \bibinfo{person}{Scott~Mayer McKinney}, \bibinfo{person}{Dean Carignan}, {and} \bibinfo{person}{Eric Horvitz}.} \bibinfo{year}{2023}\natexlab{}.
\newblock \showarticletitle{Capabilities of gpt-4 on medical challenge problems}.
\newblock \bibinfo{journal}{\emph{arXiv preprint arXiv:2303.13375}} (\bibinfo{year}{2023}).
\newblock


\bibitem[Oberoi et~al\mbox{.}(2016)]%
        {oberoi2016understanding}
\bibfield{author}{\bibinfo{person}{Simmi Oberoi}, \bibinfo{person}{Neha Chaudhary}, \bibinfo{person}{Siriesha Patnaik}, {and} \bibinfo{person}{Amarjit Singh}.} \bibinfo{year}{2016}\natexlab{}.
\newblock \showarticletitle{Understanding health seeking behavior}.
\newblock \bibinfo{journal}{\emph{Journal of family medicine and primary care}} \bibinfo{volume}{5}, \bibinfo{number}{2} (\bibinfo{year}{2016}), \bibinfo{pages}{463--464}.
\newblock


\bibitem[Parasuraman and Riley(1997)]%
        {parasuraman1997humans}
\bibfield{author}{\bibinfo{person}{Raja Parasuraman} {and} \bibinfo{person}{Victor Riley}.} \bibinfo{year}{1997}\natexlab{}.
\newblock \showarticletitle{Humans and automation: Use, misuse, disuse, abuse}.
\newblock \bibinfo{journal}{\emph{Human factors}} \bibinfo{volume}{39}, \bibinfo{number}{2} (\bibinfo{year}{1997}), \bibinfo{pages}{230--253}.
\newblock


\bibitem[Pfohl et~al\mbox{.}(2024)]%
        {pfohl2024toolbox}
\bibfield{author}{\bibinfo{person}{Stephen~R Pfohl}, \bibinfo{person}{Heather Cole-Lewis}, \bibinfo{person}{Rory Sayres}, \bibinfo{person}{Darlene Neal}, \bibinfo{person}{Mercy Asiedu}, \bibinfo{person}{Awa Dieng}, \bibinfo{person}{Nenad Tomasev}, \bibinfo{person}{Qazi~Mamunur Rashid}, \bibinfo{person}{Shekoofeh Azizi}, \bibinfo{person}{Negar Rostamzadeh}, {et~al\mbox{.}}} \bibinfo{year}{2024}\natexlab{}.
\newblock \showarticletitle{A toolbox for surfacing health equity harms and biases in large language models}.
\newblock \bibinfo{journal}{\emph{Nature Medicine}} \bibinfo{volume}{30}, \bibinfo{number}{12} (\bibinfo{year}{2024}), \bibinfo{pages}{3590--3600}.
\newblock


\bibitem[Reiley(2025)]%
        {Reiley2025}
\bibfield{author}{\bibinfo{person}{Laura Reiley}.} \bibinfo{year}{2025}\natexlab{}.
\newblock \showarticletitle{What My Daughter Told ChatGPT Before She Took Her Life}.
\newblock \bibinfo{journal}{\emph{The New York Times}} (\bibinfo{date}{18 Aug.} \bibinfo{year}{2025}).
\newblock
\urldef\tempurl%
\url{https://www.nytimes.com/2025/08/18/opinion/chat-gpt-mental-health-suicide.html}
\showURL{%
\tempurl}


\bibitem[Saab et~al\mbox{.}(2025)]%
        {saab2025advancing}
\bibfield{author}{\bibinfo{person}{Khaled Saab}, \bibinfo{person}{Jan Freyberg}, \bibinfo{person}{Chunjong Park}, \bibinfo{person}{Tim Strother}, \bibinfo{person}{Yong Cheng}, \bibinfo{person}{Wei-Hung Weng}, \bibinfo{person}{David~GT Barrett}, \bibinfo{person}{David Stutz}, \bibinfo{person}{Nenad Tomasev}, \bibinfo{person}{Anil Palepu}, {et~al\mbox{.}}} \bibinfo{year}{2025}\natexlab{}.
\newblock \showarticletitle{Advancing Conversational Diagnostic AI with Multimodal Reasoning}.
\newblock \bibinfo{journal}{\emph{arXiv preprint arXiv:2505.04653}} (\bibinfo{year}{2025}).
\newblock


\bibitem[Sayres et~al\mbox{.}(2025)]%
        {sayres2025navigating}
\bibfield{author}{\bibinfo{person}{Rory Sayres}, \bibinfo{person}{Anna Devon-Sand}, \bibinfo{person}{Mike Schaekermann}, \bibinfo{person}{Margaret~Ann Smith}, \bibinfo{person}{Patricia Strachan}, \bibinfo{person}{Grace Hong}, \bibinfo{person}{Trinh Nguyen}, \bibinfo{person}{Doris Wong}, \bibinfo{person}{Naama Hammel}, \bibinfo{person}{Dawn Siegel}, {et~al\mbox{.}}} \bibinfo{year}{2025}\natexlab{}.
\newblock \showarticletitle{Navigating Skin Concerns with AI: A Human-Centered Investigation of a Dermatology App in a Diverse Community}. In \bibinfo{booktitle}{\emph{Proceedings of the Extended Abstracts of the CHI Conference on Human Factors in Computing Systems}}. \bibinfo{pages}{1--16}.
\newblock


\bibitem[Sayres et~al\mbox{.}(2019)]%
        {sayres2019using}
\bibfield{author}{\bibinfo{person}{Rory Sayres}, \bibinfo{person}{Ankur Taly}, \bibinfo{person}{Ehsan Rahimy}, \bibinfo{person}{Katy Blumer}, \bibinfo{person}{David Coz}, \bibinfo{person}{Naama Hammel}, \bibinfo{person}{Jonathan Krause}, \bibinfo{person}{Arunachalam Narayanaswamy}, \bibinfo{person}{Zahra Rastegar}, \bibinfo{person}{Derek Wu}, {et~al\mbox{.}}} \bibinfo{year}{2019}\natexlab{}.
\newblock \showarticletitle{Using a deep learning algorithm and integrated gradients explanation to assist grading for diabetic retinopathy}.
\newblock \bibinfo{journal}{\emph{Ophthalmology}} \bibinfo{volume}{126}, \bibinfo{number}{4} (\bibinfo{year}{2019}), \bibinfo{pages}{552--564}.
\newblock


\bibitem[Shahsavar and Choudhury(2023)]%
        {shahsavar2023role}
\bibfield{author}{\bibinfo{person}{Yeganeh Shahsavar} {and} \bibinfo{person}{Avishek Choudhury}.} \bibinfo{year}{2023}\natexlab{}.
\newblock \showarticletitle{The role of AI chatbots in healthcare: a study on user intentions to utilize ChatGPT for self-diagnosis}.
\newblock \bibinfo{journal}{\emph{JMIR Preprints}} (\bibinfo{year}{2023}).
\newblock


\bibitem[Singhal et~al\mbox{.}(2023)]%
        {singhal2023large}
\bibfield{author}{\bibinfo{person}{Karan Singhal}, \bibinfo{person}{Shekoofeh Azizi}, \bibinfo{person}{Tao Tu}, \bibinfo{person}{S~Sara Mahdavi}, \bibinfo{person}{Jason Wei}, \bibinfo{person}{Hyung~Won Chung}, \bibinfo{person}{Nathan Scales}, \bibinfo{person}{Ajay Tanwani}, \bibinfo{person}{Heather Cole-Lewis}, \bibinfo{person}{Stephen Pfohl}, {et~al\mbox{.}}} \bibinfo{year}{2023}\natexlab{}.
\newblock \showarticletitle{Large language models encode clinical knowledge}.
\newblock \bibinfo{journal}{\emph{Nature}} \bibinfo{volume}{620}, \bibinfo{number}{7972} (\bibinfo{year}{2023}), \bibinfo{pages}{172--180}.
\newblock


\bibitem[Singhal et~al\mbox{.}(2025)]%
        {singhal2025toward}
\bibfield{author}{\bibinfo{person}{Karan Singhal}, \bibinfo{person}{Tao Tu}, \bibinfo{person}{Juraj Gottweis}, \bibinfo{person}{Rory Sayres}, \bibinfo{person}{Ellery Wulczyn}, \bibinfo{person}{Mohamed Amin}, \bibinfo{person}{Le Hou}, \bibinfo{person}{Kevin Clark}, \bibinfo{person}{Stephen~R Pfohl}, \bibinfo{person}{Heather Cole-Lewis}, {et~al\mbox{.}}} \bibinfo{year}{2025}\natexlab{}.
\newblock \showarticletitle{Toward expert-level medical question answering with large language models}.
\newblock \bibinfo{journal}{\emph{Nature Medicine}} \bibinfo{volume}{31}, \bibinfo{number}{3} (\bibinfo{year}{2025}), \bibinfo{pages}{943--950}.
\newblock


\bibitem[Smak~Gregoor et~al\mbox{.}(2023)]%
        {smak2023artificial}
\bibfield{author}{\bibinfo{person}{Anna~M Smak~Gregoor}, \bibinfo{person}{Tobias~E Sangers}, \bibinfo{person}{Lytske~J Bakker}, \bibinfo{person}{Loes Hollestein}, \bibinfo{person}{Carin~A Uyl-de Groot}, \bibinfo{person}{Tamar Nijsten}, {and} \bibinfo{person}{Marlies Wakkee}.} \bibinfo{year}{2023}\natexlab{}.
\newblock \showarticletitle{An artificial intelligence based app for skin cancer detection evaluated in a population based setting}.
\newblock \bibinfo{journal}{\emph{NPJ digital medicine}} \bibinfo{volume}{6}, \bibinfo{number}{1} (\bibinfo{year}{2023}), \bibinfo{pages}{90}.
\newblock


\bibitem[Sun et~al\mbox{.}(2019)]%
        {sun2019compare}
\bibfield{author}{\bibinfo{person}{Simeng Sun}, \bibinfo{person}{Ori Shapira}, \bibinfo{person}{Ido Dagan}, {and} \bibinfo{person}{Ani Nenkova}.} \bibinfo{year}{2019}\natexlab{}.
\newblock \showarticletitle{How to compare summarizers without target length? pitfalls, solutions and re-examination of the neural summarization literature}. In \bibinfo{booktitle}{\emph{Proceedings of the Workshop on Methods for Optimizing and Evaluating Neural Language Generation}}. \bibinfo{pages}{21--29}.
\newblock


\bibitem[Sutton et~al\mbox{.}(2020)]%
        {sutton2020overview}
\bibfield{author}{\bibinfo{person}{Reed~T Sutton}, \bibinfo{person}{David Pincock}, \bibinfo{person}{Daniel~C Baumgart}, \bibinfo{person}{Daniel~C Sadowski}, \bibinfo{person}{Richard~N Fedorak}, {and} \bibinfo{person}{Karen~I Kroeker}.} \bibinfo{year}{2020}\natexlab{}.
\newblock \showarticletitle{An overview of clinical decision support systems: benefits, risks, and strategies for success}.
\newblock \bibinfo{journal}{\emph{NPJ digital medicine}} \bibinfo{volume}{3}, \bibinfo{number}{1} (\bibinfo{year}{2020}), \bibinfo{pages}{17}.
\newblock


\bibitem[Tschandl et~al\mbox{.}(2020)]%
        {tschandl2020human}
\bibfield{author}{\bibinfo{person}{Philipp Tschandl}, \bibinfo{person}{Christoph Rinner}, \bibinfo{person}{Zoe Apalla}, \bibinfo{person}{Giuseppe Argenziano}, \bibinfo{person}{Noel Codella}, \bibinfo{person}{Allan Halpern}, \bibinfo{person}{Monika Janda}, \bibinfo{person}{Aimilios Lallas}, \bibinfo{person}{Caterina Longo}, \bibinfo{person}{Josep Malvehy}, {et~al\mbox{.}}} \bibinfo{year}{2020}\natexlab{}.
\newblock \showarticletitle{Human--computer collaboration for skin cancer recognition}.
\newblock \bibinfo{journal}{\emph{Nature medicine}} \bibinfo{volume}{26}, \bibinfo{number}{8} (\bibinfo{year}{2020}), \bibinfo{pages}{1229--1234}.
\newblock


\bibitem[Tu et~al\mbox{.}(2025)]%
        {tu2025towards}
\bibfield{author}{\bibinfo{person}{Tao Tu}, \bibinfo{person}{Mike Schaekermann}, \bibinfo{person}{Anil Palepu}, \bibinfo{person}{Khaled Saab}, \bibinfo{person}{Jan Freyberg}, \bibinfo{person}{Ryutaro Tanno}, \bibinfo{person}{Amy Wang}, \bibinfo{person}{Brenna Li}, \bibinfo{person}{Mohamed Amin}, \bibinfo{person}{Yong Cheng}, {et~al\mbox{.}}} \bibinfo{year}{2025}\natexlab{}.
\newblock \showarticletitle{Towards conversational diagnostic artificial intelligence}.
\newblock \bibinfo{journal}{\emph{Nature}} (\bibinfo{year}{2025}), \bibinfo{pages}{1--9}.
\newblock


\bibitem[Wang et~al\mbox{.}(2020)]%
        {wang2020human}
\bibfield{author}{\bibinfo{person}{Dakuo Wang}, \bibinfo{person}{Elizabeth Churchill}, \bibinfo{person}{Pattie Maes}, \bibinfo{person}{Xiangmin Fan}, \bibinfo{person}{Ben Shneiderman}, \bibinfo{person}{Yuanchun Shi}, {and} \bibinfo{person}{Qianying Wang}.} \bibinfo{year}{2020}\natexlab{}.
\newblock \showarticletitle{From human-human collaboration to Human-AI collaboration: Designing AI systems that can work together with people}. In \bibinfo{booktitle}{\emph{Extended abstracts of the 2020 CHI conference on human factors in computing systems}}. \bibinfo{pages}{1--6}.
\newblock


\bibitem[Wang and Cohen(2023)]%
        {wang2023health}
\bibfield{author}{\bibinfo{person}{Xun Wang} {and} \bibinfo{person}{Robin~A Cohen}.} \bibinfo{year}{2023}\natexlab{}.
\newblock \bibinfo{booktitle}{\emph{Health information technology use among adults: United States, July--December 2022}}.
\newblock \bibinfo{publisher}{US Department of Health and Human Services, Centers for Disease Control and Prevention}.
\newblock


\bibitem[Wu et~al\mbox{.}(2025)]%
        {wu2025trust}
\bibfield{author}{\bibinfo{person}{Yuheng Wu}, \bibinfo{person}{Yujie Dong}, \bibinfo{person}{Yi Mou}, {and} \bibinfo{person}{Ki~Joon Kim}.} \bibinfo{year}{2025}\natexlab{}.
\newblock \showarticletitle{How the Algorithmic Transparency of Search Engines Influences Health Anxiety: The Mediating Effects of Trust in Online Health Information Search}. In \bibinfo{booktitle}{\emph{Proceedings of the 2025 CHI Conference on Human Factors in Computing Systems}} \emph{(\bibinfo{series}{CHI '25})}. \bibinfo{publisher}{Association for Computing Machinery}, \bibinfo{address}{New York, NY, USA}, Article \bibinfo{articleno}{307}, \bibinfo{numpages}{10}~pages.
\newblock
\showISBNx{9798400713941}
\href{https://doi.org/10.1145/3706598.3713199}{doi:\nolinkurl{10.1145/3706598.3713199}}


\bibitem[Xu et~al\mbox{.}(2023)]%
        {xu2023critical}
\bibfield{author}{\bibinfo{person}{Fangyuan Xu}, \bibinfo{person}{Yixiao Song}, \bibinfo{person}{Mohit Iyyer}, {and} \bibinfo{person}{Eunsol Choi}.} \bibinfo{year}{2023}\natexlab{}.
\newblock \showarticletitle{A critical evaluation of evaluations for long-form question answering}.
\newblock \bibinfo{journal}{\emph{arXiv preprint arXiv:2305.18201}} (\bibinfo{year}{2023}).
\newblock


\bibitem[You et~al\mbox{.}(2023)]%
        {you2023beyond}
\bibfield{author}{\bibinfo{person}{Yue You}, \bibinfo{person}{Chun-Hua Tsai}, \bibinfo{person}{Yao Li}, \bibinfo{person}{Fenglong Ma}, \bibinfo{person}{Christopher Heron}, {and} \bibinfo{person}{Xinning Gui}.} \bibinfo{year}{2023}\natexlab{}.
\newblock \showarticletitle{Beyond self-diagnosis: how a chatbot-based symptom checker should respond}.
\newblock \bibinfo{journal}{\emph{ACM Transactions on Computer-Human Interaction}} \bibinfo{volume}{30}, \bibinfo{number}{4} (\bibinfo{year}{2023}), \bibinfo{pages}{1--44}.
\newblock


\bibitem[Yun and Bickmore(2025)]%
        {yun2025framing}
\bibfield{author}{\bibinfo{person}{Hye~Sun Yun} {and} \bibinfo{person}{Timothy Bickmore}.} \bibinfo{year}{2025}\natexlab{}.
\newblock \showarticletitle{Framing Health Information: The Impact of Search Methods and Source Types on User Trust and Satisfaction in the Age of LLMs}. In \bibinfo{booktitle}{\emph{Proceedings of the Extended Abstracts of the CHI Conference on Human Factors in Computing Systems}} \emph{(\bibinfo{series}{CHI EA '25})}. \bibinfo{publisher}{Association for Computing Machinery}, \bibinfo{address}{New York, NY, USA}, Article \bibinfo{articleno}{286}, \bibinfo{numpages}{7}~pages.
\newblock
\showISBNx{9798400713958}
\href{https://doi.org/10.1145/3706599.3720239}{doi:\nolinkurl{10.1145/3706599.3720239}}


\end{thebibliography}

\clearpage
\appendix
\newpage\newpage

\setcounter{figure}{0}
\makeatletter 
\renewcommand{\thefigure}{S\@arabic\c@figure}
\makeatother

\setcounter{table}{0}
\makeatletter 
\renewcommand{\thetable}{S\@arabic\c@table}
\makeatother

\section{Sample Chat History}

\begin{longtable}{c l p{0.8\textwidth}}

\toprule
\textbf{Turn} & \textbf{Speaker} & \textbf{Conversation Context} \\
\midrule
\endfirsthead

\multicolumn{3}{c}{{\bfseries \tablename\ \thetable{} -- continued from previous page}} \\
\toprule
\textbf{Turn} & \textbf{Speaker} & \textbf{Conversation Context} \\
\midrule
\endhead

\midrule
\multicolumn{3}{r}{{Continued on next page}} \\
\endfoot

\bottomrule
\endlastfoot


1 & User & > Normal acceptable range of platelets \\
\midrule
\addlinespace 

2 & Response & Hello! To help me understand your situation, I need to ask some questions. The more information you provide, the better I can pinpoint the cause of your symptoms. I will also provide my current best guess after each response, along with recommended next steps. \newline... \newline To help me understand why you're asking about platelet ranges, could you tell me: \newline\newline 1. \textcolor{blue}{Have you had a recent blood test that showed your platelet count? If so, what was the result?}\newline 2. Are you experiencing any symptoms that might be related to a low or high platelet count (e.g., easy bruising, bleeding, fatigue)? \newline 3. Do you have any known medical conditions or are you taking any medications that could affect your platelet count? \\
\addlinespace
& User & > Yes, platelet count was 642,000. \\
\midrule

3 & Response & Okay, a platelet count of 642,000 is above the normal range. This is referred to as thrombocytosis. Here are some possible causes, ranked by how well they might explain an isolated high platelet count: \newline...\newline **Questions to improve my answer:** \newline 1. \textcolor{blue}{**Have you had any recent infections, surgeries, or injuries?**} \newline 2. **Do you have any known inflammatory conditions (e.g., rheumatoid arthritis, inflammatory bowel disease) or iron deficiency?** \newline 3. **Are you experiencing any unusual bleeding or bruising, or any symptoms suggestive of blood clots?** \\
\addlinespace
& User & > No recent infections, surgeries, or injuries. \\
\midrule

4 & Response & Okay, that helps narrow things down. The absence of recent infections, surgeries, or injuries makes reactive thrombocytosis less likely but doesn't rule it out entirely... Given your elevated platelet count of 642,000 and the absence of recent triggers, here's an updated list of possible causes:\newline ...\newline **Questions to improve my answer:** \newline 1. \textcolor{blue}{**Do you have any known inflammatory conditions (e.g., rheumatoid arthritis, inflammatory bowel disease) or iron deficiency?**} \newline 2. **Are you experiencing any unusual bleeding or bruising, or any symptoms suggestive of blood clots?** \newline 3. **Have you had a complete blood count (CBC) done recently, or just the platelet count? If a CBC was done, what were your red blood cell and white blood cell counts?** \\
\addlinespace
& User & > No known inflammatory conditions. \\
\midrule

5 & Response & Okay, that's helpful. Knowing that you don't have any *known* inflammatory conditions helps to slightly decrease the likelihood of reactive thrombocytosis due to a diagnosed inflammatory disorder...\newline ...\newline **Questions to improve my answer:** \newline 1. **Are you experiencing any unusual bleeding or bruising, or any symptoms suggestive of blood clots?** \newline 2. **Have you had a complete blood count (CBC) done recently, or just the platelet count? If a CBC was done, what were your red blood cell and white blood cell counts?** \newline 3. \textcolor{blue}{**Are you taking any medications or supplements (including over-the-counter)?**} \\
\addlinespace
& User & > Taking hydroxurea once a day. \\
\midrule

6 & Chatbot & Okay, that's a very important piece of information. \newline 
...\newline 
**Questions to improve my answer:**
\newline 
1. **Have you been officially diagnosed with Essential Thrombocythemia (ET) or another Myeloproliferative Neoplasm (MPN)?**  \newline 
2. \textcolor{blue}{**When was your last appointment with your hematologist, and what did they say about your platelet count at that time?** } \newline 
3. **Are you experiencing any new or worsening symptoms since starting hydroxyurea?** \\ 
\addlinespace
 & User & > At my last appointment with my hematologist, platelet count was 425,000. \\
\midrule

7 & Chatbot & Okay, that's helpful context. Knowing that your platelet count was 425,000 at your last hematologist appointment, and is now 642,000, suggests a recent upward trend despite being on hydroxyurea. This reinforces the need to investigate the cause of the increase.\newline
...\newline
**To provide more targeted information, it would be helpful to know:**
\newline
1. **Have you been officially diagnosed with Essential Thrombocythemia (ET) or another Myeloproliferative Neoplasm (MPN)?**  \newline
2. **Are you experiencing any new or worsening symptoms since starting hydroxyurea, or any symptoms related to blood clots or bleeding?**  \newline
3. \textcolor{blue}{**When was your last appointment with your hematologist, besides the one where your platelet count was 425,000? Was a complete blood count performed and were there any other abnormal results?**} \\
\addlinespace

 & User & > My platelet count was 642,000, now it's gone down to 425,000. \\
\bottomrule

\caption{\textbf{Sample Multi-turn Chat History (P10).} This example demonstrates an interaction with a deferred-answer prototype. The blue highlighting indicates the AI's question that the user’s answer directly corresponds to.}

\label{tab:chat-history}

\end{longtable}

\section{Supplemental Methods}

\subsection{Agents used throughout the study}

Across the 4 studies in this paper, we tested a range of conversational agents that were iteratively updated in terms of both model behavior and user interface. The full set of variants we tested is described in Table \ref{tab:agents_summary}. For the first study, we also include an illustration of the different conversational dynamics between the Immediate response behavior used by Agent A1 (as well as all other agents in later studies) and the Delayed-answer behavior used by Agent A2.

\begin{table}[!ht]
    \centering
    \begin{tabular}{m{0.5cm}|m{4.8cm}|m{5cm}|m{2.8cm}}
    \toprule[2pt]
    \textbf{AI} & \textbf{Brief description} & \textbf{AI Details} & \textbf{Study Examining AI} \\
    \midrule[2pt]
1a & Context-seeking + Immediate answer & Initial model prompted to aid cause-of-symptom-intent (COSI) questions. Provides set of matching conditions followed by clarifying questions & Study 1 (qualitative) \\
    \midrule
1b & Context-seeking + Delayed answer  & Variant of AI, with system prompt to only ask clarifying questions for at least 4 (but no more than 10) turns, in order to make the set of matching conditions as relevant as possibly & Study 1 (qualitative) \\
    \midrule[1.5pt]
2a & Generalized context-seeking & Refined version of A1, provides initial answer but tuned to address both COSI and NCOSI questions & Study 2 (qualitative) \\
    \midrule
2b &  Baseline model for 2a & Base model for Wayfinding AI 2a, without any additional tuning or system prompts to engage in context-seeking & Study 2 (qualitative) \\
    \midrule[1.5pt]
3a & Wayfinding AI candidate, 1-column & Refined version of 2a, with clarifying questions consistently formatted at end of each prompt and presented underneath user prompts in a single column & Study 3 (qualitative) \\
    \midrule
3b & Wayfinding AI candidate, 2-column & Refined version of 2a, in a 2 column format: clarifying questions underneath user prompts in left column, answers in right column. Answers are initially collapsed and can be expanded by users selecting expandable chips in left column. & Study 3 (qualitative) \\
    \midrule[1.5pt]
4a & Wayfinding AI candidate & Refined version of 3b, using a 2-column format but where answers are expanded by default and always immediately visible & Study 4 (quantitative) \\
    \midrule
4b & Baseline model for 4a & Baseline model for Wayfinding AI 4a, without any additional tuning or system prompts to engage in context-seeking & Study 4 (quantitative) \\
    \bottomrule[2pt]
    \end{tabular}
\caption{Summary of different AI versions tested in this study.}
\label{tab:agents_summary}
\end{table}

\begin{figure*}[!htp]
    \centering
    \includegraphics[width=\textwidth]{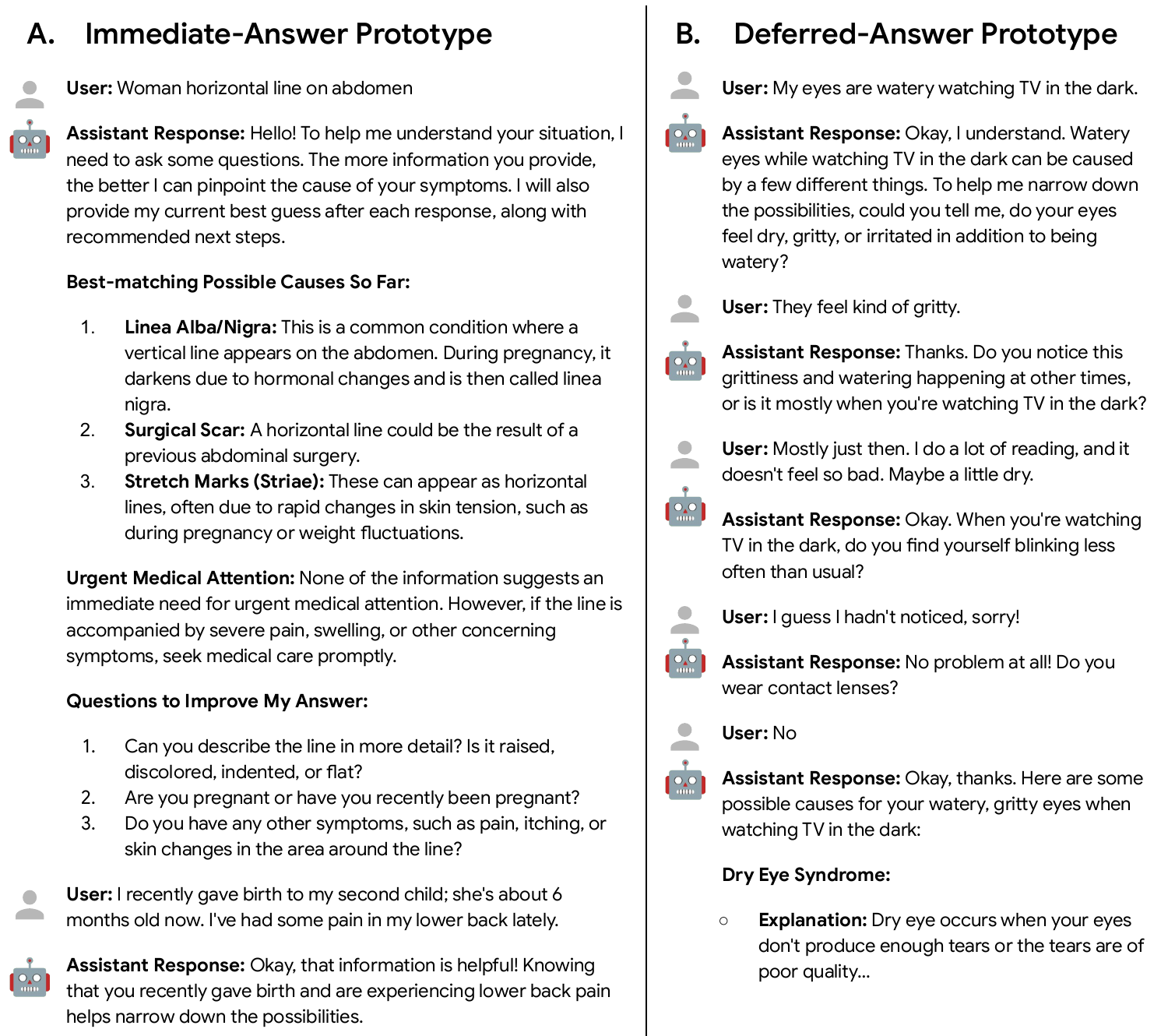}
    \caption{Illustration of differences in agent behavior tested in Study 1. Left, AI 1a provided immediate answers to responses, with context-seeking clarifying questions at the bottom of each response. Right, AI 1b deferred answers for 4 or more conversation turns, instead only asking clarifying questions. During this section of the session, the researcher engaged in a remote moderated usability session with the participant, understanding whether the two behaviors met or differed from their expectations for a conversational AI.}
    \label{fig:study_1_agents}
    \Description{This illustrates two mock conversations in Study 1. In the left, an AI chabot provides immediate answers to a user prompt, while in the right, the AI first asks clarifying questions, only providing an answer after several turns.}
\end{figure*}

\newpage
\clearpage

\subsection{Conversation classification}

We used a large language model (Gemini 2.5 Pro) to classify conversations, as well as individual conversational turns.

\subsubsection{Cause-of-symptom intent classification}

For identifying cause-of-symptom-intent from non-cause-of-symptom-intent, we used the following prompt:

\begin{verbatim}    
You are an expert health assistant.

Your task is to analyze a conversation between a User and an AI Assistant,
and determine if it reflects an intent related to discovering the cause of
somebody's health symptoms (either the user or someone else).

The conversation will be provided below. User's comments are provided after
the string "USER:", and the Assistant's replies are after the string
"ASSISTANT:". Sometimes, the Assistant response will include markup like
"(RIGHT)", "(LEFT)", or "(CHIP)"; this reflects formatting of the answer
in an interface. Ignore it for determining the intent of the conversation.

Return 'COSI' if the conversation has cause of symptom intent, and 'NCOSI' if
it does not. If the intent can not be reasonably determined, return 'Unsure'.
Do not return any response other than 'COSI', 'NCOSI', or 'Unsure'.

CONVERSATION:
\end{verbatim}

\subsubsection{User turn type classification}

For classifying conversation turn types, we used the following prompt:

\begin{verbatim}
You are an expert in analyzing conversational user behavior. You will be provided with a 
user's previous turn query, assistant's response and the user's subsequent response if the 
user responded. Your task is to classify the user's response to the assistant's previous 
turn in a multi-turn conversation. Categorize the user response into **one and only one** of 
the following categories:

1. RESPONSE_WITH_INFO - User responds with information solicited by the assistant.
    *   Signals: Responding with facts that were asked by the assistant.
    *   Examples: "my height is 5ft", "no I do not currently take any medications", "this 
    has been happening for the last 6 months"

2. REJECTION - User indicates the response is incorrect, irrelevant, misleading, or
fundamentally flawed. Includes direct disagreement, challenges to facts, and statements of 
misinformation.
    *   Signals: "That's wrong," "This is incorrect," "You're lying," "I don't agree," 
    "That's not what I asked," challenges with evidence, skeptical questions implying 
    disbelief if the intent is clearly rejection.
    *   Examples: "That's not the right answer." "This is incorrect, it was 1945 not 1946." 
    "You're confusing two different things." "Are you sure that's correct? Because I found 
    X."

3.VERIFICATION_VALIDATION - User expresses uncertainty or seeks confirmation of the 
information provided, **without asserting it's wrong**. Asks for evidence, justification, or
source information to **assess reliability**. This category applies when the user is unsure 
and seeks to confirm rather than directly disagreeing.
    *   Signals: "Are you sure?", "How do you know?", "Can you provide a source?", "What's 
    the basis for that?" **without directly contradicting the assistant**.
    *   Examples: "Are you sure about that number? Can you show me where you got that 
    from?", "Where did you get that information?", "What makes that source reliable?  Is it 
    peer-reviewed?"

4. REFINEMENT_CLARIFICATION - User provides more details to clarify the original query, 
narrow the scope, or add constraints to improve relevance. The focus is on making the 
assistant's subsequent response *more useful*, not questioning its validity.
    *   Signals: Adding keywords, specifying criteria, excluding terms, defining terms.
    *   Examples: "I meant Windows 8 **on a tablet**." "Exclude results from Wikipedia." 
    "What about vegan options?"

5. ELABORATION - User requests more detail on a specific aspect of the response that the 
assistant already provided. The user isn't questioning the response itself, just wanting 
more information *on the topics already introduced*. This also includes cases when the user
wants to see a different viewpoint or consider another factor, alternative solutions **to 
the problem or situation that the assistant has addressed**.
    *   Signals: "Tell me more about...", "Explain X in more detail," "Give me an example 
    of...", "What are the benefits/drawbacks **of what you just said**?"
    *   Examples: "Explain how that works **that you just mentioned**." "Can you give me 
    some real-world examples **of this approach**?" "What are the specific advantages of 
    this approach **you're describing**?"

6. RELATED_TOPIC_EXPLORATION - User asks about related but not directly implied topics, 
seeking a broader understanding **that goes beyond the original scope AND beyond the details 
already offered by the assistant**. This is a "tangential but related" question.
    *   Signals: Questions about related concepts, tangential connections, and similar ideas
    **not already discussed**.

    *   Examples: (After travel planning) "What are some good bars nearby **the hotel you 
    recommended**?" (After discussing a historical event) "What were the social consequences 
    of that **war you just discussed**?"

7. COMPARATIVE -  User wants a comparison between different options **presented in the 
response**. *This only applies if the assistant has already offered multiple options*.
    *   Signals: How does X compare to Y?, Pros and cons, which is better for me, compare
    and contrast.
    *   Examples: "What are the pros and cons of **each of those options**?", "How does 
    option A compare to option B? **You mentioned both**", "Which one is better for X 
    purpose? **You gave me 3 options**"

8. TASK_ORIENTED_FOLLOWUP - User asks for assistance with a task that logically follows from 
the initial response. This is about getting the assistant to *do something* based on the 
information it provided, rather than just asking questions.
    *   Signals: Requests for automation, generation, translation, coding examples, shopping
    lists, etc.
    *   Examples: (After finding a recipe) "Can you create a shopping list **for that 
    recipe**?" (After finding information on a software library) "Can you show me a code 
    example **using that library**?"

9. PIVOT - User abruptly changes the topic to something unrelated to the previous discussion.
    *   Signals: Introducing a completely new subject with no obvious connection to the 
    prior conversation.
    *   Examples: (After discussing travel) "By the way, do you know anything about quantum
    physics?"

10. CONVERSATION_CLOSURE - User signals the end of the conversation, with no need for 
followup
    *   Signals: Signaling end of conversation by saying thank you, or stop or related 
    phrases.
    *   Examples: "thank you!", "thanks", "ty", "that answered my question", "that's all",
    etc.

11. NO_RESPONSE - the user did not respond to the assistant's message.

**Instructions:**

1.  You are given a previous turn user query, assistant response, and next turn user 
response (None if the user did not respond/speak).
2.  Determine the type of the next turn user response based on the categories above. Note 
that the categories include cases where the user responded with statements, as well as cases
where the user responded with a question or questions. Based on the type of response, 
determine the best category.
3.  If the user did not respond, output NO_RESPONSE
4.  Explain your reasoning in one or two sentences, justifying why you chose a specific 
category. This should refer to the definitions provided.
5.  Choose only one category that best describes the user response behavior to the assistant.

Output your response in the format:

Reason: {{reason}}
ResponseType: {{user response type}}

**Input:**
\end{verbatim}

\subsubsection{AI question-type classification}
For classifying AI responses as containing clarifying questions, open-ended questions, or no questions, we used the following prompt:

\begin{verbatim}
You are an expert conversation analyst. Your task is to classify the assistant's
response provided below.

# ASSISTANT RESPONSE
{assistant_response}

# TASK
Analyze the "ASSISTANT RESPONSE" above. Choose ONLY ONE of the following categories that best describes the
Assistant's Response:
- 'Clarifying Question': The assistant asked for more detail.
- 'Open-ended Question': The assistant asked a broad, engaging question.
- 'No Question': The assistant did not ask a follow-up question.

Your output must be ONLY the category name (e.g., 'Clarifying Question').
\end{verbatim}

\section{Supplemental Results}

\begin{table} 
    \begin{tabular}{lcc}
    \toprule[2pt]
    \textbf{Breakdown}& \textbf{\# of Participants} & \textbf{Percent (\%)} \\

    \midrule[2pt]
    \multicolumn{1}{m{9cm}}{\textbf{Age Range}} &  &  \\
    21 to 30 & 19 & 14.62 \\
    31 to 40 & 23 & 17.69 \\
    41 to 50 & 35 & 26.92 \\
    51 to 60 & 34 & 26.15 \\
    61 to 70 & 7 & 5.38 \\
    71 or over & 1 & 0.77 \\
    n/a & 11 & 8.46 \\
    
    \midrule
    \multicolumn{1}{m{9cm}}{\textbf{Gender}} &  &  \\
    Man & 73 & 56.15 \\
    Woman & 57 & 43.85 \\
    
    \midrule    
    \multicolumn{1}{m{9cm}}{\textbf{Racial or ethnic groups}} &  &  \\
    Asian & 11 & 8.46 \\
    Black or African American & 13 & 10.00 \\
    Hispanic, Latino or Spanish Origin & 16 & 12.31 \\
    White & 93 & 71.54 \\

    \midrule
    \multicolumn{1}{m{9cm}}{\textbf{Completed highest degree}} &  &  \\
    Less than a high school degree & 1 & 0.77 \\
    High school diploma or the equivalent (for example: GED) & 12 & 9.23 \\
    Some college credit, no degree & 31 & 23.85 \\
    Associates degree (for example: AA, AS) & 7 & 5.38 \\
    Bachelor’s degree (for example: BA, BS) & 50 & 38.46 \\
    Master’s degree (for example: MA, MS, MEng, MEd, MSW, MBA) & 24 & 18.46 \\
    Doctorate degree (for example: PhD, EdD) & 5 & 3.85 \\

    \midrule    
    \multicolumn{1}{m{9cm}}{\textbf{Confidence in filling out medical forms by myself (without asking for help from someone else)}} &  &  \\
    Extremely confident & 79 & 60.77 \\
    Very confident & 29 & 22.31 \\
    Moderately confident & 16 & 12.31 \\
    Slightly confident & 6 & 4.62 \\

    \midrule    
    \multicolumn{1}{m{9cm}}{\textbf{Confidence in using mobile phone apps by myself (without asking for help from someone else)}} &  &  \\
    Extremely confident & 95 & 73.08 \\
    Very confident & 21 & 16.15 \\
    Moderately confident & 10 & 7.69 \\
    Slightly confident & 3 & 2.31 \\
    Not at all confident & 1 & 0.77 \\

    \midrule    
    \multicolumn{1}{m{9cm}}{\textbf{Frequency in using AI chatbots (e.g. ChatGPT, Claude, Gemini)}} &  &  \\
    Most days & 24 & 18.46 \\
    Several times a week & 32 & 24.62 \\
    Several times a month & 24 & 18.46 \\
    Several times a year & 31 & 23.85 \\
    Never. I have never used an AI chatbot or don’t know what it is. & 19 & 14.62 \\
    \bottomrule[2pt]
    Total & 130 & 100 \\
    \bottomrule[2pt]
    \end{tabular}

    \caption{Participant metadata summary for Study 4.}
    \label{tab:study_4_participants}
\end{table}

\begin{figure*}[ht]
    \centering
    \includegraphics[width=300pt]{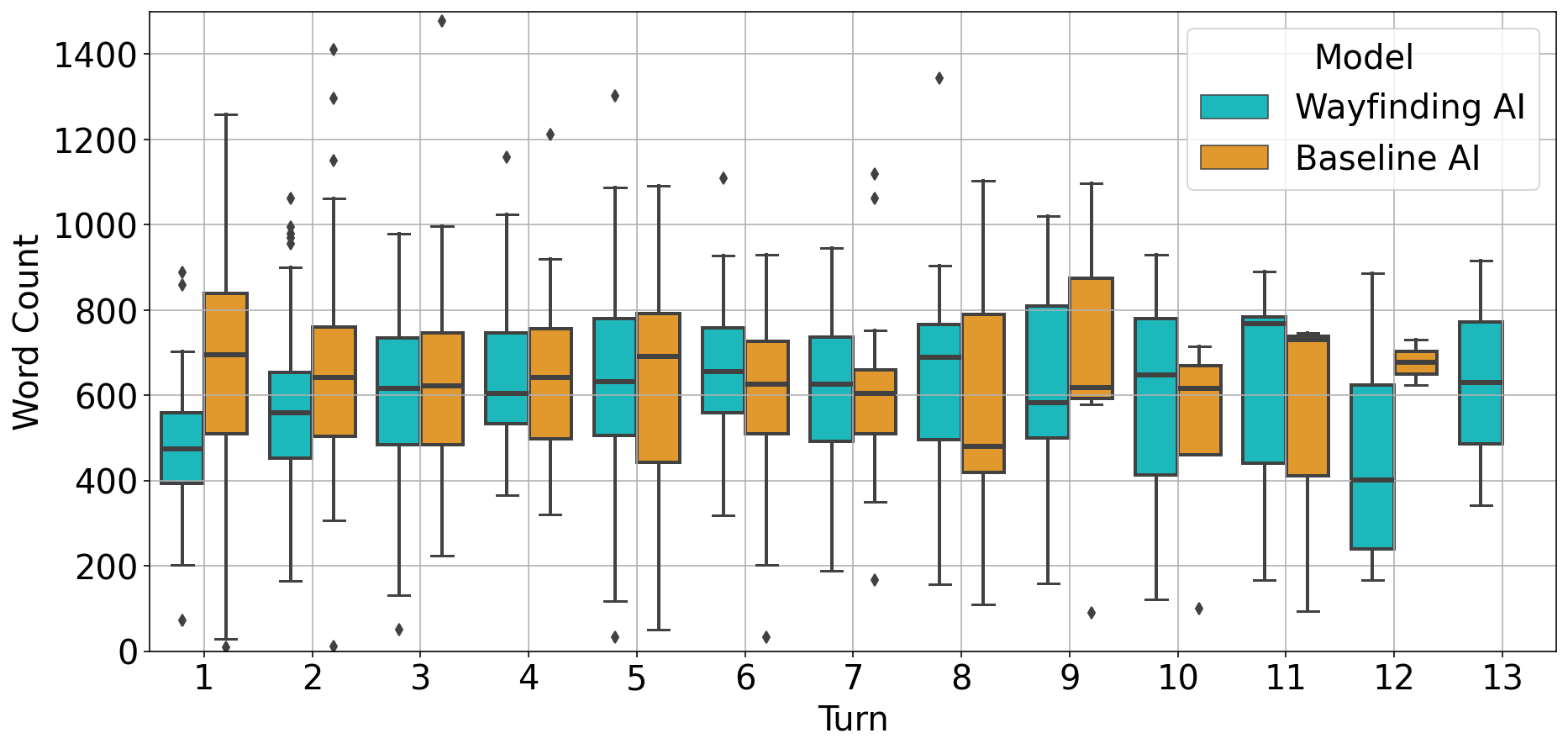}
    \caption{Comparisons of each turn's word counts between Wayfinding AI and Baseline AI.}
    \label{fig:wordcount_summary}
    \Description{This figure shows box plots of word counts for each conversational turn for the Wayfinding AI and Baseline AI responses. The word counts for the Baseline AI responses tend to be higher for the first several turns.}
\end{figure*}

\begin{figure*}[ht]
    \centering
    \includegraphics[width=\textwidth]{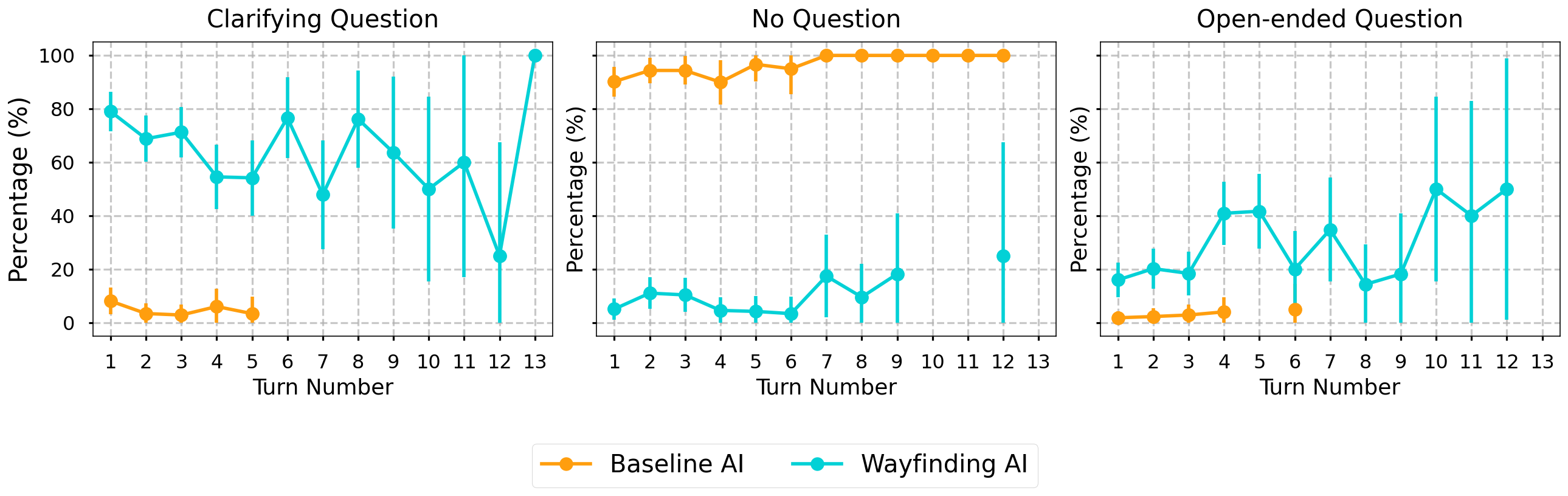}
    \caption{Rates of AI responses containing Clarifying questions (left), No questions (center), and Open-ended questions (right) over conversational turns, for the Baseline AI (orange) and Wayfinding AI (blue).}
    \label{fig:chatbot_response_type}
    \Description{This figure shows graphs of the percentage of each AI's responses in three different types: No questions, clarifying questions, and open-ended questions. Each graph shows the conversation turn number on the X axis, with the percentage of conversations containing that type of question on the Y axis. Separate lines show the rates for the Wayfinding AI and Baseline AI. The Wayfinding AI has consistently high rates of Clarifying questions, while the Baseline AI has few to none. The Baseline AI has very high rates of asking No questions, while the rate is below 10\% for the Wayfinding AI. Both AIs have fairly low rates of Open-ended questions, though it is somewhat higher for the Wayfinding AI, between 10 and 40\%.}
\end{figure*}

\end{document}